\documentclass[]{article}
\usepackage{emulateapj}
\usepackage{graphicx}

\advance \voffset by -.50cm\relax

\def\msun{{\rm ~M}_{\odot}}
\def\rsun{{\rm ~R}_{\odot}}

\def\mpy{{\rm ~M}_{\odot} {\rm ~yr}^{-1}}

\begin{document}

\title{Black Hole Spin Evolution: Implications for Short-hard Gamma Ray Bursts
and Gravitational Wave Detection}

 \author{Krzysztof Belczynski\altaffilmark{1,2,3}, 
         Ronald E.\ Taam\altaffilmark{4,5}, 
         Emmanouela Rantsiou\altaffilmark{4}, 
         Marc van der Sluys\altaffilmark{4}}

 \affil{
     $^{1}$ Los Alamos National Laboratory, 
            CCS2/ISR1 Group, P.O. Box 1663, MS D466,
            Los Alamos, NM 87545\\
     $^{2}$ Oppenheimer Fellow\\
     $^{3}$ New Mexico State University, Dept of Astronomy,
            1320 Frenger Mall, Las Cruces, NM 88003\\
     $^{4}$ Northwestern University, Dept of Physics \& Astronomy,
            2145 Sheridan Rd, Evanston, IL 60208\\
     $^{5}$ ASIAA/National Tsing Hua University - TIARA, Hsinchu, Taiwan\\
     kbelczyn@nmsu.edu, r-taam, emmanouela, sluys@northwestern.edu}

 \begin{abstract}
The evolution of the spin and tilt of black holes in compact black hole 
- neutron star and black hole - black hole binary systems is investigated
within the framework of the coalescing compact star binary model for short gamma ray 
bursts via the population synthesis method. Based on recent results on accretion 
at super critical rates in slim disk models, estimates of natal kicks, 
and the results regarding fallback in supernova models, we obtain the 
black hole spin and misalignment. It is found that the spin parameter,
$a_{\rm spin}$, is less than 0.5 for initially non rotating black holes and 
the tilt angle, $i_{\rm tilt}$, is less than $45^\circ$ for 50\% of the systems in 
black hole - neutron star binaries. Upon comparison with the results 
of black hole - neutron star merger calculations we estimate that 
only a small fraction ($\sim 0.01$) of these systems can lead to the formation 
of a torus surrounding the coalesced binary potentially producing  
a short-hard gamma ray burst.  On the other hand, for high initial black hole spin 
parameters ($a_{\rm spin}>0.6$) this fraction can be significant ($\sim 0.4$). 
It is found that the predicted gravitational radiation signal for our simulated 
population does not significantly differ from that for non rotating black holes. 
Due to the {\rm (i)} insensitivity of signal detection techniques to the black 
hole spin and the {\rm (ii)} predicted overall low contribution of black hole binaries 
to the signal we find that the detection of gravitational waves are not greatly 
inhibited by current searches with non spinning templates. It is pointed out that 
the detection of a black hole - black hole binary inspiral system with LIGO or 
VIRGO may provide a direct measurement of the initial spin of a black hole.  
\end{abstract}


\section{Introduction}

Double compact object binaries have attracted much attention in recent years 
as primary sources of gravitational wave radiation (GR) and as potential 
sites for the short-hard gamma ray burst (GRB) phenomenon. As such, intensive
searches for the inspiral signal from double neutron stars (NS-NS), double 
black holes (BH-BH) and mixed black hole - neutron star (BH-NS) systems are 
currently underway at ground based GR observatories (e.g., LIGO or VIRGO; 
for a recent review see Kalogera et al. 2007).  Of these systems, the mergers 
of NS-NS and BH-NS systems have been suggested to produce short-hard GRBs (for 
a recent review see Nakar 2007).  However as new NS-NS binaries are discovered 
(e.g., Lorimer 2005), there has yet to be a 
detection of a BH-NS or BH-BH system. 

Despite the accumulation of observational data on short-hard GRBs from the 
HETE-II and SWIFT satellites, the origin of these GRBs still remains elusive. 
Theoretical predictions of the compact merger model have been compared to 
observations in the hope of identifying the possible progenitor (e.g., Nakar, 
Gal-Yam \& Fox 2005; Belczynski et al. 2006), but some specifics of the model 
are still lacking. For example, to assess the validity of BH-NS merger as a 
short-hard GRB progenitor, it is usually assumed that all these mergers produce 
a GRB. However, recent hydrodynamical simulations (e.g., Setiawan, Ruffert \& 
Janka 2004; Faber et al. 2006; Shibata \& Uryu 2006; Rantsiou et al. 2007)
indicate that only a fraction of these  binaries can potentially produce 
a GRB. Specifically, the existence of a thick torus of material is required such 
that it can be rapidly accreted onto the central BH to produce a GRB. To satisfy 
this constraint, the NS must be disrupted prior to its final plunge below the BH 
event horizon with sufficient angular momentum such that the remnant material can 
form an accretion torus. In this case, only compact systems with specific parameters 
(spin, mass ratio, and BH spin tilt with respect to the orbital angular momentum 
axis) can provide the requisite initial conditions for GRB production. Accordingly, 
we report on the results of our investigation of the spin magnitude and tilt in 
merging BH-NS binaries to estimate the fraction of BH-NS systems that can potentially 
produce short-hard GRBs. 

The spin of the BH in BH-BH and BH-NS binaries is of special interest in the search 
for gravitational wave signatures in existing data streams from ground-based 
observatories. On the one hand, population synthesis studies (e.g., Fryer, Woosley \& 
Hartmann 1999; Nelemans, Yungelson \& Portegies Zwart 2001; Belczynski et al. 2007) 
attempt to provide realistic merger rates and the characteristic properties of the 
merging binaries, while, on the other, detailed general relativistic  
calculations of mergers (e.g., Baker et al. 2006; Buonanno, Cook \& Pretorius 2007) 
attempt to predict 
the exact shape of the GR signal. These latter studies are essential for guiding 
the search for the inspiral signal that currently involves the use of a 
limited number of pre-calculated gravitational wave templates (e.g., Abbott 
et al. 2005;  Abbott et al. 2006). It is well known that if the spin of a BH is 
significant and if it is misaligned with respect to the orbital angular momentum
axis, the shape of GR signal will differ drastically from the non-spinning/aligned 
case (Apostolatos et al. 1994; Kidder 1995; see also \S\,2.5).  Yet, so far the 
search methods presented in the literature (e.g., Abbott et al. 2005; Abbott et 
al. 2006) employ non-spinning templates only.  

The previous studies of BH spins in compact binaries in the context of BH-NS 
systems were carried out by Kalogera (2000), Grandclement et al. (2004) and 
O'Shaughnessy et al.\ (2006). The spin up of the BH via binary evolutionary 
processes was estimated, and it was found that accretion in the common envelope 
(CE) phase was a major contributing factor in significantly spinning up the BHs. 
These early results are reassessed and extended here since {\em (i)} now only 
very few BH-BH progenitor systems are predicted to evolve through the CE phase 
(Belczynski et al. 2007), {\em (ii)} for systems evolving through the CE phase 
the Bondi-Hoyle accretion mode may overestimate the accretion (and spin-up) rates 
(see \S\,2.2), and {\em (iii)} during the stable mass transfer phases the degree 
of spin up requires reevaluation in view of the possibility that the BH can 
accept mass at rates significantly exceeding the Eddington accretion rate (e.g., 
Abramowicz et al. 1988; Ohsuga et al. 2005; Ohsuga 2007).  Since the 
predictions of compact object spin misalignment depend critically on the natal 
kick distribution, we make use of the most recent work on natal kicks by Hobbs et 
al. (2005).  Finally, we also extend previous studies to include the double black 
hole binary population.  

In the next section, the various elements of our model are described. The 
results of our BH spin calculations are presented in \S\,3. Finally, in 
\S\,4 we conclude and discuss the implications of our findings.  

\section{Model}

\subsection{Population Synthesis Model}

Binary population synthesis is used to calculate the populations of close 
BH-NS and BH-BH binaries that merge within 10 Gyr. The formation of double 
compact objects is modeled via binary evolutionary processes in the absence 
of stellar dynamical processes (e.g., such as in the cores of globular clusters). 
The formation of BH-BH systems in dense environments was recently studied by 
Portegies Zwart \& McMillan (2000), O'Leary et al. (2006) and Sadowski et
al. (2008). No relevant studies are available for BH-NS binaries, as their 
formation rate in dense environments is likely negligible or small, as was 
predicted for NS-NS systems (e.g. Phinney 1991; Grindlay, Portegies Zwart \& 
McMillan 2006).  

Our population synthesis code, {\tt StarTrack},  was initially developed to 
study double compact object mergers in the context of GRB progenitors
(Belczynski, Bulik \& Rudak 2002b) and gravitational-wave inspiral sources 
(Belczynski, Kalogera, \& Bulik 2002a: hereinafter BKB02). In recent years {\tt 
StarTrack} has undergone major updates and revisions in the physical treatment 
of various binary evolution phases, and especially the mass transfer phases. 
The new version has already been tested and calibrated against observations and 
detailed binary mass transfer calculations (Belczynski et al.\ 2008a), and has 
been used in various applications (e.g., Belczynski \& Taam 2004; Belczynski et 
al.\ 2004; Belczynski, Bulik \& Ruiter 2005; Belczynski et al. 2006; Belczynski 
et al.\ 2007). The physics updates that are most important for compact object 
formation and evolution include: a full numerical approach for the orbital evolution 
due to tidal interactions, calibrated using high mass X-ray binaries and open 
cluster observations, a detailed treatment of mass transfer episodes fully 
calibrated against detailed calculations with a stellar evolution code, updated 
stellar winds for massive stars, and the latest determination of the natal
kick velocity distribution for neutron stars (Hobbs et al.\ 2005).  For helium 
star evolution, which is of a crucial importance for the formation of double 
neutron star binaries (e.g., Ivanova et al.\ 2003; Dewi \& Pols 2003), we 
have applied a treatment matching closely the results of detailed evolutionary 
calculations.  If the helium star fills its Roche lobe, the systems are examined 
for the potential development of a dynamical instability, in which case they 
are evolved through a CE phase, otherwise a highly non-conservative mass
transfer ensues. We treat CE events using the energy formalism (Webbink 1984),
where the binding energy of the envelope is determined from the set of He star 
models calculated with the detailed evolutionary code by Ivanova et al.\ (2003). 
In case the CE is initiated by a star crossing the Hertzsprung gap (HG) we 
assume a merger and abort further binary evolution. This is due to the fact 
that there is no clear core-envelope boundary (and no entropy jump as for 
more evolved stars) in the interior structure of HG donors to facilitate the
formation of a remnant binary system. As a consequence, a large decrease in 
the formation efficiency of close double compact binaries results (Belczynski 
et al. 2007).  For a detailed description of the revised code we refer the 
reader to Belczynski et al.\ (2008a).

Since the study of Belczynski et al. (2007) there was an update of
rejuvenation treatment for main sequence stars, and the method presented by
Tout et al. (1997) was introduced as described in detail in Belczynski et
al. (2008a; see their \S\,5.6). The resulting changes for close BH-NS
binaries are negligible. For close BH-BH binaries we note $\sim 40\%$ drop 
in coalescence rates, a drop that is insignificant compared to over 2 
orders of magnitude uncertainty in the model rates (Belczynski et al. 2007).

\subsection{BH Accretion Model: Spin Magnitude}

The pioneering studies of accretion onto a BH began with the seminal 
work of Shakura \& Sunyaev (1973) and Thorne (1974). 
In this study, we employ the formalism presented by Brown et al. (2000) to 
describe the spin evolution of a BH as taken from the energy and angular momentum 
derived from the Killing 
vector of the Kerr metric (Boyer \& Lindquist 1967). The Boyer-Lindquist coordinates 
allow for a continuous transformation across the double horizons of the Kerr metric, 
preserving the essential singularity at $r=0$, and also allowing us to address the 
stress-energy tensor inside the BH.  As such, the angular momentum is calculated for 
the entire mass of the BH resulting in the following evolution equations. 

The BH spin parameter is defined as 
\begin{equation}
a_{\rm spin} = {J c \over M_{\rm bh}^2 G}
\label{mod01}
\end{equation}
where $M_{\rm bh}$ denotes the BH mass, $J$ its angular momentum, and $G$ and $c$ 
are the gravitational constant and the speed of light, respectively. The angular 
momentum, $l$, and energy, $E$, of the accreted material with rest mass 
$M_{\rm rest}$ can be expressed as 

\noindent
\begin{eqnarray}
l= & \left[{R_{\rm lso}^2 - a_{\rm sp} \sqrt{2 R_{\rm bh} R_{\rm lso}} + a_{\rm sp}^2 
     \over  R_{\rm lso} (R_{\rm lso}^2-{3 \over 2} R_{\rm bh} R_{\rm lso} + a_{\rm sp}
     \sqrt{2 R_{\rm bh} R_{\rm lso}})^{1/2}} \right] \nonumber \\
   & \times c \sqrt{{R_{\rm bh} R_{\rm lso} \over 2}} M_{\rm rest} 
\label{mod02}
\end{eqnarray}

\noindent
\begin{eqnarray}
E= & \left[{R_{\rm lso}^2 - R_{\rm bh} R_{\rm lso} + a_{\rm sp} \sqrt{R_{\rm bh} R_{\rm lso}/2} 
     \over  R_{\rm lso} (R_{\rm lso}^2-{3 \over 2} R_{\rm bh} R_{\rm lso} + a_{\rm sp}
     \sqrt{2 R_{\rm bh} R_{\rm lso}})^{1/2}} \right] \nonumber \\
   & \times c^2 M_{\rm rest}
\label{mod03}
\end{eqnarray}
where $R_{\rm bh}=2GM_{\rm bh}/c^2$ is the Schwarzschild radius of a BH, 
$a_{\rm sp}=J/M_{\rm bh}c = a_{\rm spin} (GM_{\rm bh}/c^2)$, and the last stable orbit
radius is calculated from 
\noindent
\begin{equation}
R_{\rm lso}= {R_{\rm bh} \over 2} \{3+z_2-[(3-z_1)(3+z_1+2 z_2)]^{1/2}\} 
\label{mod04}
\end{equation}
with
\noindent
\begin{equation}
z_1= 1+\left(1- {4 a_{\rm sp}^2 \over R_{\rm bh}^2}\right)^{1/3} 
\left[ \left(1 + {2a_{\rm sp} \over R_{\rm bh}} \right)^{1/3} + 
\left(1 - {2a_{\rm sp} \over R_{\rm bh}} \right)^{1/3} \right] \nonumber \\
\label{mod05}
\end{equation}
and 
\noindent
\begin{equation}
z_2= \left( 3 {4 a_{\rm sp}^2 \over  R_{\rm bh}^2} + z_1^2  \right)^{1/2}.   
\label{mod06}
\end{equation}

The accretion of $M_{\rm rest}$ onto a BH changes its gravitational mass to 
\begin{equation}
M_{\rm bh,f}=M_{\rm bh,i} + {E \over c^2}
\label{mod07}
\end{equation}
where the indices $i,f$ correspond to the initial (pre-) and final (post-accretion)
values. The accretion of $M_{\rm rest}$ onto a BH changes its angular momentum
to 
\begin{equation}
J_{\rm f}=J_{\rm i} + l
\label{mod08}
\end{equation}
where the initial angular momentum is obtained from eq.~\ref{mod01} 
$J_{\rm i}=a_{\rm spin,i} M_{\rm bh,i}^2 G/c$, and the new BH spin is
calculated from
\begin{equation}
a_{\rm spin,f} = {J_{\rm f} c \over M_{\rm bh,f}^2 G}.
\label{mod09}
\end{equation}
Since the BH spins at formation are unknown, we perform our calculations for 
a wide range of the initial values of spin parameter, including the non-spinning 
BH case ($a_{\rm spin,init}=0$) as well as rapidly rotating BH cases
($a_{\rm spin,init}>0.9$). We refer to case of $a_{\rm spin,init}=0$ as low spin, 
$a_{\rm spin,init}=0.55$ as moderate spin, and $a_{\rm spin,init}=0.9$ as high spin.

Given the mass transfer rates obtained from the population synthesis, we make 
use of calculations of super-critical accretion flows around BHs to estimate 
the mass accretion rate (e.g., Abramowicz et al. 1988; Ohsuga et al. 2002; Ohsuga 
et al. 2005; Ohsuga 2007).  Recently, Ohsuga (2007) demonstrated that photon 
trapping in slim accretion disk models is important, allowing accretion at rates 
significantly exceeding the critical value ($\dot M_{\rm crit}$, see below). 
In particular, for a flow rate (e.g., transfer rate from a donor star 
$\dot M_{\rm don}$) of $\sim 1000 \times \dot M_{\rm crit}$, matter was accreted at 
a rate at the level of $\dot M_{\rm acc} \gtrsim 100 \times \dot M_{\rm crit}$, with 
the remaining matter lost in disk outflow due to strong radiation pressure effects.  
For lower flow rates ($\dot M_{\rm don} \lesssim 100 \times \dot M_{\rm crit}$), the disk 
undergoes limit cycle oscillations.
We fit\footnote{The mass accretion rates onto a BH were presented only
graphically in Ohsuga (2007).} the recent results presented by Ohsuga (2007; see his 
Fig.~4) to obtain a prescription for the mass accretion rate onto a BH as
follows\\
$\log \left( {\dot M_{\rm acc} \over \dot M_{\rm crit}} \right)  =$\\
\begin{equation}
 \left\{ \begin{array}{ll}
  \log \left( {|\dot M_{\rm don}| \over \dot M_{\rm crit}} \right) & 
     |\dot M_{\rm don}| \leq \dot M_{\rm crit}\\
  0.544 \log \left( {|\dot M_{\rm don}| \over \dot M_{\rm crit}} \right) & 
     \dot M_{\rm crit} < |\dot M_{\rm don}| \leq 10 \times \dot M_{\rm crit}\\
  0.934 \log \left( {|\dot M_{\rm don}| \over \dot M_{\rm crit}} \right) - 0.380 & 
     |\dot M_{\rm don}| > 10 \times \dot M_{\rm crit}\\
\end{array}
\right.
\label{mod10}
\end{equation}
where we adopt $\dot M_{\rm crit}=2.6 \times 10^{-8} (M_{\rm bh}/10 \msun)
\mpy$ from Ohsuga (2007)\footnote{Ohsuga (2007) expresses critical mass flow 
rate in terms of $L_{\rm edd}/c^{2}$, where $L_{\rm edd}$ is the critical
Eddington luminosity and $c$ is the speed of light.}. At the lowest 
mass transfer rates $\dot M_{\rm don} \lesssim \dot M_{\rm crit}$, it is 
assumed that all the matter transferred is accreted by the black hole. 
It is noted that the accretion limits adopted from Ohsuga (2007) are
based not on full GR calculations, but on calculations performed using 
a gravitational potential for non-rotating black holes developed by Paczynsky \& Wiita
(1980). We have also adopted the Ohsuga (2007) results for the viscosity parameter 
of $\alpha=0.5$.  The introduction of black hole spin in the Ohsuga (2007) 
calculations would tend to decrease the limiting accretion rate. The effect 
would be especially important for rapidly rotating black holes. Therefore, our results 
provide upper limits on accretion and mass accumulation on black holes in binary systems. 

In the case of a NS accretor the accretion is limited to the critical Eddington 
accretion rate. The critical Eddington rate for NS is taken to be 
$1.7 \times 10^{-8} \mpy$ in case of hydrogen accretion and $2.9 \times 10^{-8} \mpy$ 
in case of heavier element accretion. 

The above prescriptions (BH and NS accretion) are adopted for the case of 
dynamically stable Roche lobe overflow (RLOF). The transferred 
material that is not accreted onto the compact object is assumed to be  
ejected from the system, carrying away the specific orbital angular momentum of 
the compact object. The subsequent change in the binary orbit is readily obtained 
(e.g., Belczynski et al. 2008a).

In the case of mass transfer proceeding on a dynamical timescale, a CE phase 
will develop.  The amount of mass accreted during this phase, denoted as 
$ \Delta M_{\rm acc}$, is taken to be given by 
\begin{equation}
\Delta M_{\rm acc}= f_{\rm ce} \times \Delta M_{\rm bondi} 
\label{mod11}
\end{equation}
where $\Delta M_{\rm bondi}$ is the amount of mass accreted if accretion 
proceeds at the Bondi-Hoyle rate, and $f_{\rm ce}=0.1$ is a scaling factor
adopted for both neutron star and black hole accretors (e.g., Ricker \& Taam
2008). 
Since the matter within an accretion cylinder is characterized by density 
and velocity gradients, accretion is not spherically symmetric.  
The numerical calculations of Ruffert (1999) confirm that the 
accretion rate can be significantly lower than the Bondi-Hoyle rate. 
We use a numerical approach, presented in 
Belczynski et al. (2002, see their Appendix), to estimate $\Delta M_{\rm bondi}$. 
For a NS accretor, the amount of mass accreted is sufficient to mildly recycle the 
pulsar (e.g., Zdunik, Haensel \& Gourgoulhon 2002; Jacoby et al. 2005). This 
particular approach results in a surprisingly good match between the observed 
and predicted masses of pulsars in Galactic double neutron star systems (Belczynski 
et al. 2008b). As only part of the donor envelope is accreted onto the 
compact object, the remainder is ejected at the expense of the change in orbital 
energy. We use the standard energy balance (e.g., Webbink 1984) in which fully efficient 
energy transfer from the orbit to the envelope ($\alpha_{\rm ce}=1$) is assumed to 
calculate the change of orbital separation. 

\begin{figure}
\includegraphics[width=1.0\columnwidth]{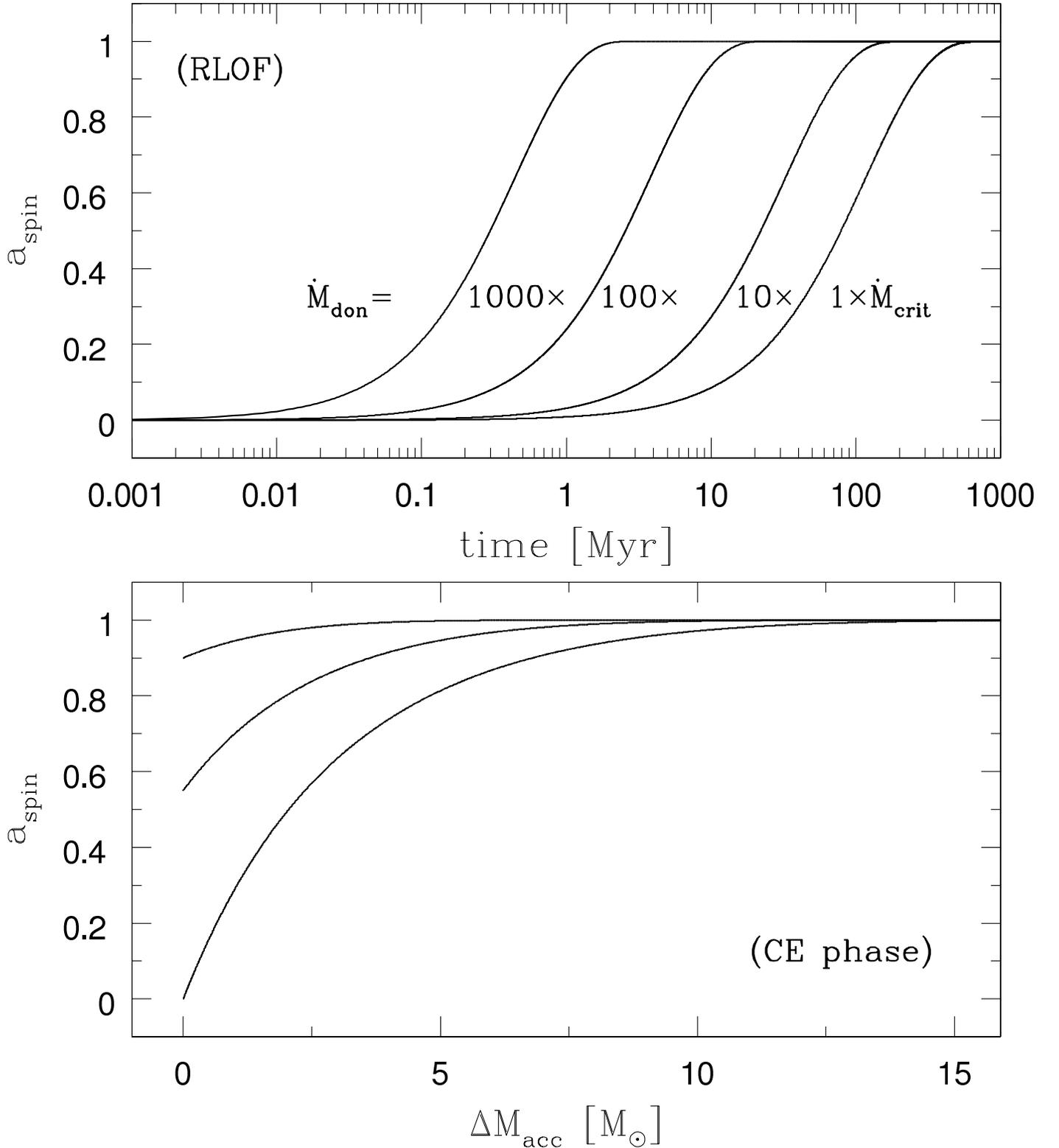}
\caption{
Spin evolution of a $10 \msun$ black hole. 
Top panel: spin up dependence on mass transfer rate. Note that there is a
need for a prolonged RLOF phase ($\gtrsim 100$ Myr, the time not available
for BH-NS progenitors) if mass transfer rate is limited or close to the 
critical ($\sim 1-10 \times \dot M_{\rm crit}$) rate, while relatively short time ($\sim
1$ Myr) is needed for significant BH spin up at very high transfer rates 
($\sim 1000 \times \dot M_{\rm crit}$). Note that not all transferred material is
accreted onto a BH (see eq.~\ref{mod10}).
Bottom panel: spin up dependence on amount of accreted (rest) mass. Note
that a large amount of mass needs to be accreted ($\gtrsim 4-7 \msun$) to
significantly spin up a BH ($a_{\rm spin} \gtrsim 0.9$) if its initial spin is
low to moderate.}
\label{model}
\end{figure}

An example of the spin magnitude evolution for a BH with initial mass of $10 \msun$ 
is illustrated in Figure~\ref{model}. In the upper panel, we consider the BH spin 
evolution for several different mass transfer rates $\dot M_{\rm don}=  1 \times,\ 
10 \times,\ 100 \times, 1000 \times \dot M_{\rm crit}$, where  $\dot M_{\rm
crit}=2.6 \times 10^{-8} \mpy$ is the critical accretion rate for a $10 \msun$ BH.  
For an initially non-spinning BH (i.e., $a_{\rm spin} = 0$), it is clear that 
a prolonged RLOF phase ($\sim 200$ Myr) is required to significantly 
spin up the BH ($a_{\rm spin} \gtrsim 0.9$) if the mass transfer proceeds at the 
critical rate ($\dot M_{\rm crit}$). If mass transfer is as high as 
$100 \times \dot M_{\rm crit}$ then a shorter time ($\sim 1 - 10$ Myr) is
required to significantly increase BH spin.  In the lower panel of Figure~\ref{model} 
we display the BH spin up as a function of accreted mass (e.g., in CE phase) 
for the three different initial BH spin values of $a_{\rm spin} = 0,\ 0.55,\ 0.9$. It 
is easily seen that if the BH initial spin is small/moderate, a significant 
amount of mass must be accreted ($\gtrsim 4-7 \msun$) to increase the spin to 
high values ($a_{\rm spin} \gtrsim 0.9$).

\subsection{Supernova Explosion Model: Spin Tilt}

To estimate the degree of misalignment (tilt) of the BH spin axis 
relative to the orbital angular momentum axis of the binary system, 
we assume no tilt initial conditions. That is, {\em (i)} the spin axes 
of both stars in the binary system have spins that are parallel to the orbital 
angular momentum axis, {\em (ii)} once a compact object is formed in a 
core collapse/supernova explosion the spin direction is preserved (i.e., 
the BH spin preserves the same direction as the spin of collapsing star). 
In this case, the tilt results from the change of the orbital 
plane due to the natal kick the compact object receives in the core 
collapse/supernova explosion. 

We assume that the distribution of natal kicks is isotropic and the magnitude 
is obtained from a single Maxwellian with $\sigma = 265$ km s$^{-1}$ (Hobbs 
et al. 2005) modified in the following way
\begin{equation}
V_{\rm kick} = (1-f_{\rm fb}) V
\label{kick01}
\end{equation}
where $V$ is the kick magnitude drawn from the Hobbs et al. (2005) distribution, 
and $f_{\rm fb}$ is a fallback parameter, i.e., the fraction (from 0 to 1) of 
the stellar envelope that falls back onto the compact object. For a NS 
compact object, no fall back is assumed (energetic SN explosion) and full 
kicks are applied ($f_{\rm fb}=0$).  On the other hand, for the most massive 
BHs, formed silently (no SN explosion) in a direct collapse ($f_{\rm fb}=1$) 
of a massive star to a BH, it is assumed that no natal kick is imparted. 
This would occur for the most massive stars (initial, zero age main 
sequence stars with masses $\gtrsim 40 \msun$: Fryer 1999; Fryer \& Kalogera 
2001) that form massive BHs ($M_{\rm bh} \gtrsim 9 \msun$; Belczynski et al. 
2008a).  Lower mass BHs are formed accompanying a SN explosion, however 
only a fraction of the progenitor envelope is ejected and the rest is retained 
by the newly formed BH. In these cases the kick is decreased by an amount 
dependent on the expected fall back of mass (for more details see Belczynski 
et al. 2008a). 

The effect of the supernova explosion on the binary orbit is followed in 
the general case of eccentric orbits.  In particular, we chose a random 
position on the orbit where the explosion takes place and use evolutionary 
formulae to estimate the mass of the compact object. The mass ejection is 
assumed to be spherical, and the expelled material is removed carrying 
the specific orbital angular momentum of the exploding star. The newly formed 
compact object receives the natal kick, changing the direction and magnitude 
of its velocity around its companion star. If the resulting orbit is unbound 
the binary evolution is terminated and the two stars are followed as single 
objects. However, for a bound orbit the orbital parameters are recalculated 
to include the inclination of the orbit. Here, the change of the orbital 
inclination is equal to the change in direction of compact object spin with 
respect to the orbital angular momentum direction.  In the case of double 
compact objects a progenitor system experiences two core collapse/supernova 
events and the respective tilts of the first and second born compact 
object are then given by:
\begin{equation}
\begin{array}{ll}
  i_{\rm tilt,1} = \Delta i_{\rm SN1} + \Delta i_{\rm SN2} & {\rm first\ born} \\
  i_{\rm tilt,2} = \Delta i_{\rm SN1} + \Delta i_{\rm SN2} & {\rm second\ born}
\end{array} 
\label{tilt1}
\end{equation} 
where $\Delta i_{\rm SN1}$ and $\Delta i_{\rm SN2}$ denote the relative change 
of the orbital inclination in a first and a second core collapse/supernova event
respectively.  In previous studies (Kalogera 2000; Grandclement et al. 2004; 
O'Shaughnessy et al. 2005) only BH-NS systems were considered and only the 
spin evolution of the BH was followed\footnote{The spin of a NS is small 
compared to that for a BH in BH-NS binaries with a massive BH [e.g., see 
Kalogera (2000) and references therein]. However, we follow the spin of both 
components since we also include BH-BH systems in this study.}. It is assumed 
that the spin of the first born compact object (BH) is aligned with orbit at the 
time of second core collapse/supernova event (i.e., $i_{\rm tilt,1} = 
\Delta i_{\rm SN2}$). This assumption is based on the fact that a mass 
transfer episode that occurs between two supernova phases, will tend to 
align the spin of a BH with orbital angular momentum axis. Alignment for the 
progenitor of second born compact object can also occur in the same manner
during a mass transfer episode between the two core collapse/supernova events 
when the progenitor filling its Roche lobe is a subject to strong tidal 
interactions. For this case, it is likely that $i_{\rm tilt,2} = \Delta 
i_{\rm SN2}$ for BH-NS systems and for some BH-BH binaries. We point out 
that this assumption may not be justified in evolutionary scenarios of 
BH-NS formation where only very little mass ($\sim 0.5 \msun$) is accreted 
onto a massive BH ($\sim 10 \msun$), or in the case of BH-BH formation where 
mass transfer does not occur between the two core collapse/supernova events 
(Belczynski et al.  2007, see their Table 1, model A).  We shall discuss 
the effect of the above assumptions on the spin tilt in the formation of double 
compact object binaries in the following sections.

\subsection{Short-hard GRB Model}

The mergers of BH-NS binaries have been proposed to give rise to short-hard 
GRBs. A common ingredient in such theoretical models is the requirement of 
a thick torus surrounding the BH.  The subsequent accretion of matter in the 
torus releases the gravitational potential energy required to power the GRB. 
Setiawan et al.(2004) suggest that $\nu\bar{\nu}$-annihilation in a 
low density funnel above the BH along its spin axis can deposit energy 
at a rate of $\sim 10^{50\,}$erg $s^{-1}$, accounting for a total energy 
release of some $10^{49}\,$erg for the case of a torus characterized by a 
mass $\gtrsim 0.1 \msun$ and a high viscosity orbiting a BH with spin $a_{\rm spin} 
\sim 0.6$.  Recent calculations of BH-NS mergers carried out 
using a general relativistic treatment by three independent groups (Shibata 
\& Uryu 2006; Faber et al. 2006; Rantsiou et al. 2007) 
have explored the outcome of a merger event as a function of mass 
ratio, equation of state for the NS, and the misalignment of BH spin with 
respect to the orbital angular momentum axis. 

For BH-NS mergers with a BH in the mass range of $\sim 3-4 \msun$, Shibata 
\& Uryu (2006) conclude that the disruption of a NS by a low mass BH will 
lead to the formation of a low mass disk ($\sim 0.1 \msun$) around the BH 
which could potentially power a short-hard GRB. The formation of a massive 
disk was not found in their simulations, leading them to conclude that 
systems with massive disks ($\sim 1 \msun$) cannot be formed in BH-NS mergers 
with a non-rotating BH. Similar results were found by Faber et al. (2006), 
who employed a fully relativistic treatment in their simulations of BH-NS 
mergers for low mass ratio ($q=0.1$) systems.  In particular, most of the 
infalling NS mass is accreted promptly onto the BH, with a 
fraction ($\sim 25\%$) of NS remaining bound in the form of a disk.

In contrast to these two groups, who assumed a non spinning BH, Rantsiou 
et al. (2007) investigated the effect of the BH spin angular 
momentum and the BH spin misalignment for BH-NS mergers with mass ratio 
$q \sim 0.1$ (i.e., for a $15 \msun$ BH). They found that both BH spin and 
their tilts play an important role in the outcome of the merger. Specifically, 
only for high BH spins ($a_{\rm spin} > 0.9$) and tilts in the range of 
$20-40^\circ$, can the merger result in the ejection of significant fraction 
(up to $40\%$) of NS mass, part of which will remain bound to form a thick 
torus of mass $\sim 0.1 \msun$ around the BH.  

\begin{figure}
\includegraphics[width=0.55\columnwidth,angle=0]{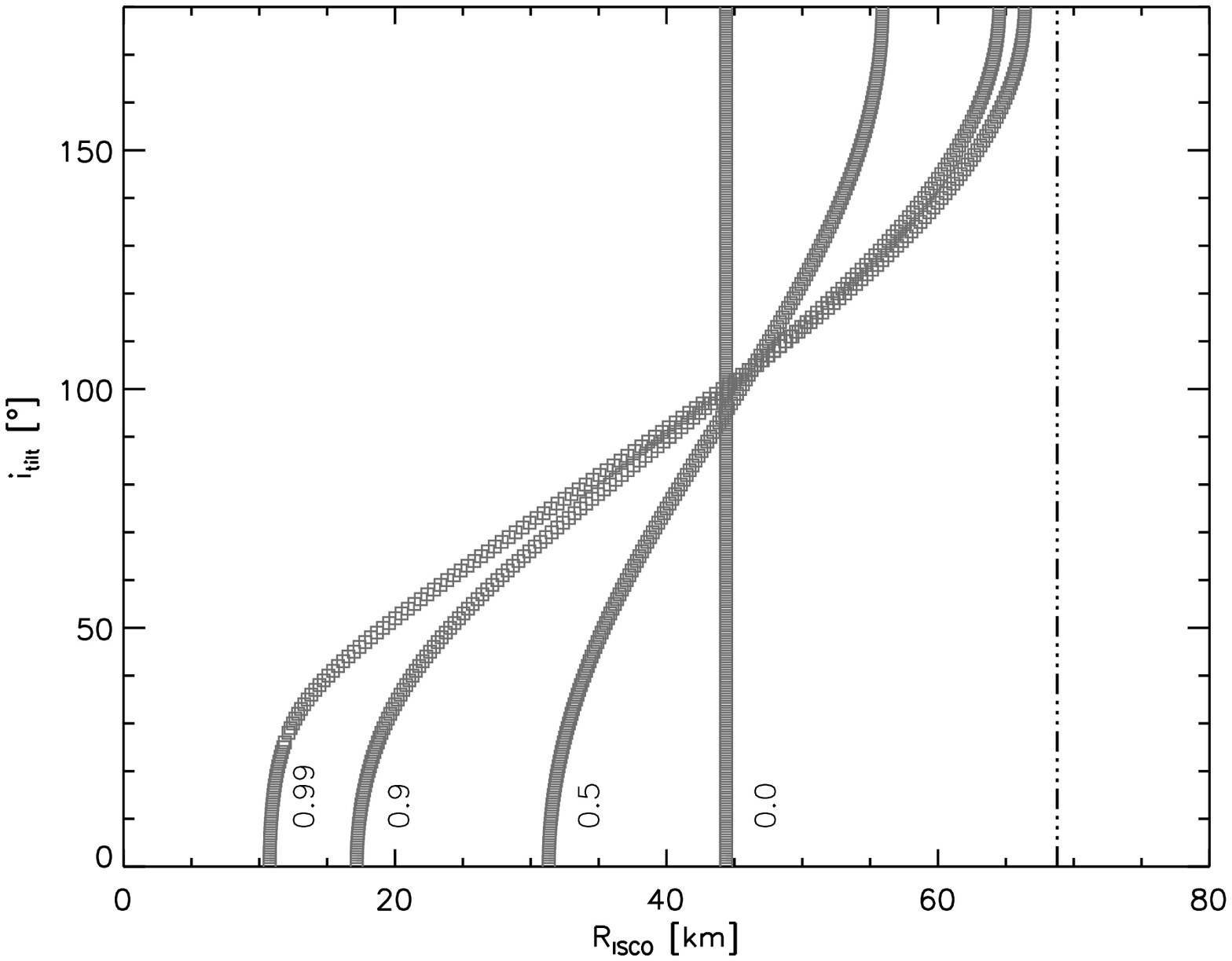}\\
\includegraphics[width=0.55\columnwidth,angle=0]{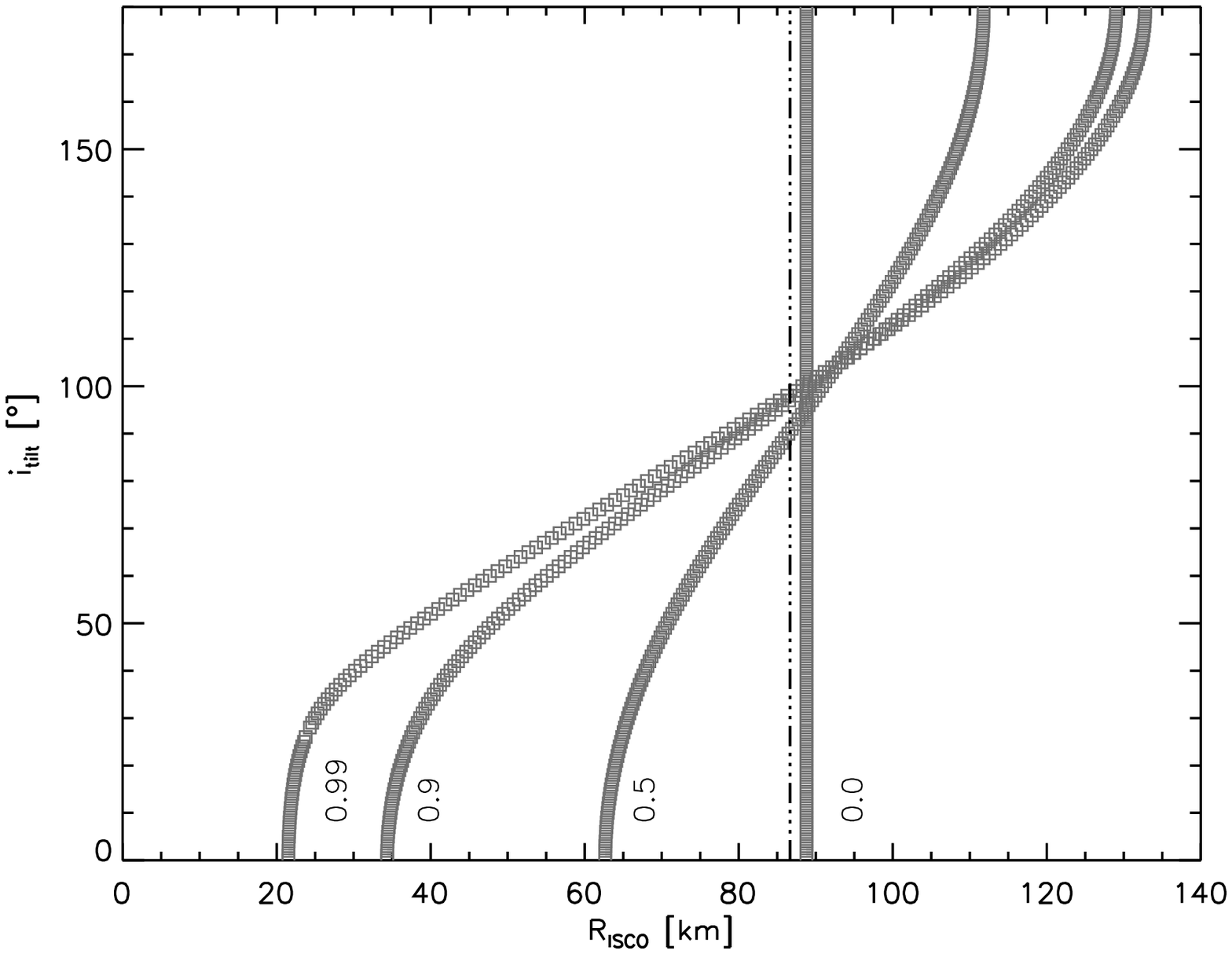}\\
\includegraphics[width=0.55\columnwidth,angle=0]{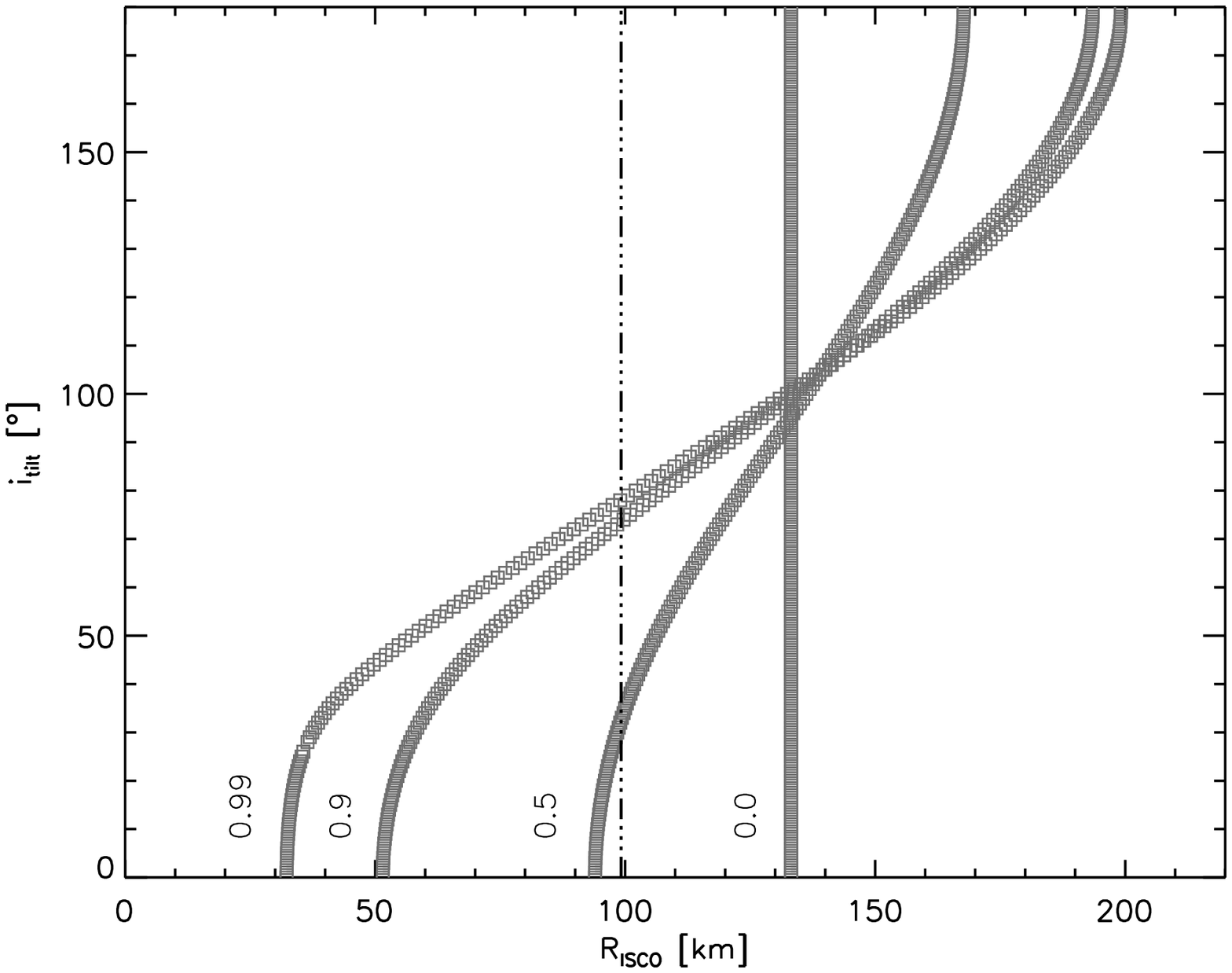}
\caption{
ISCO radius dependence on BH spin tilt (with respect to orbital angular momentum) 
for three representative BH masses: 5 (top panel), 10 (middle) and $15 \msun$ 
(bottom). The gray lines show the dependence for various BH spins ($a_{\rm spin}
=0,0.5,0.9,0.99$). The vertical dashed lines represent the disruption radius for 
a NS of $1.4 \msun$ and $R_{\rm ns}=15\,$km. Note that the disruption radius needs
to be well outside ISCO radius to provide a necessary (but not sufficient) 
condition for a GRB.}
\label{isco}
\end{figure}

A key issue for the outcome of a BH-NS merger is the relative position of 
the innermost stable circular orbit (ISCO), denoted by $R_{\rm isco}$, 
of the BH and the disruption radius (or tidal radius $R_{\rm tid}$) of the
inspiraling NS. Disruption must occur well outside the ISCO, for otherwise, 
the entire NS plunges below the event horizon of a BH, leaving no material 
to initiate a GRB. Therefore, a necessary condition for the production of 
a GRB is $R_{\rm tid} > R_{\rm isco}$. However, this is not a sufficient
condition, since {\em (i)} the disrupted material must form an accretion 
torus around BH and {\em (ii)} there must be sufficient material ($\sim 0.01-
0.3 \msun$, e.g., Ruffert \& Janka 1999) in the torus to power a GRB. 
To provide for sufficient material, $R_{\rm tid} \gtrsim 2 \times R_{\rm 
isco}$.  The high inclination orbits ($i_{\rm tilt} \gtrsim 40-90^\circ$) are 
excluded since material not accreted by the BH in these mergers is ejected
and does not form an accretion torus around the BH.  We note that the position 
of the ISCO not only depends on the BH spin, but also on its tilt 
with respect to the orbital angular momentum vector. Furthermore, the relative 
position of $R_{\rm tid}$ and $R_{\rm isco}$ depends on the mass ratio of the NS 
to BH.  Figure~\ref{isco} illustrates the dependence and provides insight on the 
binary parameters that could allow for the formation of a disk/torus around the 
BH. Based on Figure~\ref{isco} and the results of the available simulations 
(see above) for BH-NS mergers some constraints on the characteristics of 
BH-NS binaries can be placed on those mergers that could potentially power a 
short-hard GRB. Table~\ref{grb} summarizes our criteria for a short-hard GRB 
production from BH-NS mergers.

The ISCO for a particle in an orbit of inclination $i$ around a BH of mass $M$ 
and arbitrary Kerr parameter $a_{\rm sp}$ (defined after eq.3) is found by 
solving the set of equations
\begin{equation} 
R=0=R'=R''
\label{isco1}
\end{equation}
where $R$ is defined by
\begin{equation}
\Sigma^2 \left (\frac{dr}{d\tau} \right
)^2=[E(r^2+a_{\rm sp}^2)-a_{\rm sp}L_z]^2-\Delta[r^2+(L_z-a_{\rm sp}E)^2+Q]\equiv R
\label{isco2}
\end{equation}
with $ \Sigma=r^2+a_{\rm sp}^2\cos^2\theta $ and $\Delta=r^2-2Mr+a_{\rm sp}^2$
(Carter 1968; Hughes 2000).
The condition $R=0$ defines a circular orbit; an orbit of constant radius $r_0$. 
The condition on the first derivative $R'=0$ means that the particle's radial 
acceleration is also zero at $r_0$ as required for an orbit to remain circular. 
A condition on the second derivative $R''<0$ in general guarantees that
the orbit is stable (actually determining that the effective potential has a
minimum at that point) and if specifically $R''=0$ then the orbit has the 
smallest allowed radius.
In the above equation $E$ and $L_z$ are the conserved energy and conserved
angular momentum per unit rest mass respectively and $Q$ is the Carter
constant. One can numerically find the ISCO-inclination dependence for orbits 
around a BH of given $a_{\rm sp}$ by varying the orbit's inclination $i$ (defined as
$cos(i)=\frac{L_z}{\sqrt{L_z^2+Q}}$) from $0^o$ to $180^o$ and solving Eqs.
\ref{isco1} for $L_z$, $e$ and $r$.

\subsection{Gravitational Inspiral Signal}

The gravitational-wave signal resulting from an inspiral of a stellar mass 
compact object binary (BH-NS or BH-BH) can be detected by ground-based 
interferometric detectors, such as LIGO or VIRGO. These signals are strongly 
affected by the presence of non-parallel spin of the compact object with 
respect to the orbital angular momentum direction in the binary, mostly due 
to the orbital precession that is induced by the spin-orbit interaction.
In Figure~\ref{grsig} we present a comparison of the gravitational wave signals  
for spinning and non-spinning BHs. The inspiral is presented in terms of
amplitude strain $h$ (relative test mass shift) at the output of a given 
detector, defined by a linear combination of two independent polarization 
states of the gravitational wave ($h_{\rm +},\ h_{\rm x}$)  convolved with 
the interferometric antenna pattern (e.g., Apostolatos et al. 1994).  
 
Figure~\ref{grsig} shows a noiseless waveform shape in the  
time domain, as it would be detected by one of the LIGO detectors and the
VIRGO detector.  The signal is detectable once the gravitational wave 
frequency enters the detector band at 40\,Hz for LIGO or 30\,Hz for VIRGO 
until the binary reaches the ISCO. The calculation was performed for a 
binary consisting of a 10\,$M_\odot$ BH and a 1.4\,$M_\odot$ NS, at a 
distance of 30\,Mpc. For reference, the upper panel shows the signal for 
the non spinning case, while in the lower panels, the BH is characterized 
by $a_{\rm spin} = 0.1,\ 0.3$ with a misalignment angle between spin and orbital 
angular momentum taken to be $i_{\rm tilt} = 35^\circ$. Here, the NS is 
assumed to have negligible spin angular momentum. The waveforms were created 
in the 1.5 post Newtonian approximation for the phase and Newtonian amplitudes 
based on a simple precession model (Apostolatos et al. 1994) to describe the 
effect of spin.  The conversion of the global signal to the local signal for each 
detector\footnote{The detector signal that was calculated for LIGO and VIRGO  
depends (strongly) on the position in the sky, which was chosen  
randomly.} was performed using the network routines of the Monte-Carlo 
code developed by Roever, Meyer \& Christensen (2006).

\section{Results}

The evolution of massive binaries that eventually form BH-NS and 
BH-BH binary systems are followed, considering only those systems 
with coalescence times (orbital decay due to GR) less than 10 Gyr 
as potential GR or short-hard GRB sources. The updated merger rates 
of double compact objects have already been presented and discussed 
in the light of the recent input physics developments (Belczynski et al 
2007), while an initial comparison of the physical properties of NS-NS binaries
with the observed Galactic population and some implications of NS-NS 
and BH-NS mergers for short-hard GRBs are presented in Belczynski et al. 
(2008b; 2008c). Here, we discuss the potential effects of BH spin evolution 
on the theoretically expected rates of these phenomena.

\subsection{BH-NS Binaries}

In Table~\ref{bnchan} we present the accretion history of double compact 
object progenitors. The accretion phases, taking place either during the 
CE or stable RLOF phase between the two core collapse/supernova events 
(first: SN1, and second: SN2) are listed. Only the first born compact 
object (in SN1) may increase its spin magnitude during one of the accretion 
phases, while the second born compact object is not subject to accretion 
and spin evolution. The binary population synthesis results show that the 
most frequent accretion mode for the BH-NS progenitors is through the CE 
phase only ($\sim 73\%$).  Accretion via both CE and RLOF phases amounts to 
$\sim 26\%$ of the cases, while accretion through RLOF only or no accretion 
at all is rarely encountered ($\sim 1\%$).  
Note that we do not discuss here the accretion history prior to the first
compact object formation, although every accretion event in the history of
a given progenitor is taken into account in our population synthesis
calculations.     

In Figure~\ref{dMacc} we show the amount of mass accreted onto the BH 
in BH-NS progenitors. Three different accretion modes, identified in 
Table~\ref{bnchan}, are presented separately. It is clear that the amount 
of accreted mass is rather small and of the order of $\sim 0.1 \msun$ for CE
accretion, and $\sim 0.3 \msun$ for the combined CE and RLOF accretion. 
This small amount directly reflects the limits placed on the accretion in the 
CE phase, and the adopted CE efficiency ($\alpha_{\rm ce}=1$; 
see \S\,4 for discussion of this dependence) for the progenitor stars entering 
the CE phase ($\sim 15 \msun$ core helium burning donor with $\sim 10 \msun$ BH).
In the case for which accretion occurs during the RLOF phase, the donors are 
helium stars. These stars are not very massive stars ($\sim 3-6 \msun$),
and the mass transfer rate very often proceeds on a thermal timescale that 
can reach $\gtrsim 1000 \times \dot M_{\rm crit}$, however, only a small 
fraction ($\lesssim 10\%$) of transferred material is accreted onto the BH 
(see eq.~\ref{mod10}). Therefore, due to the limited mass accretion in 
both cases, the mass accreted onto the BH, in the formation of close BH-NS 
systems is not very large, thereby setting limits on the expected BH spin up 
in these binaries. We note that most of the systems evolve and accrete 
during the CE phase.  Had a full Bondi-Hoyle accretion rate been adopted, 
the accreted mass would have increased to $\sim 1 \msun$.  Even such an  
amount is insufficient to significantly increase the spin of a massive BH 
(see Fig.~1).

The distribution of final BH spins are presented in Figure~\ref{aspin} for 
the three different initial conditions: non spin ($a_{\rm spin}=0$), moderate 
spin ($a_{\rm spin}=0.55$) and high spin ($a_{\rm spin}=0.9$). It can be 
seen that, in all cases, BHs in BH-NS binaries increase their spins
through accretion. For the initially non spinning case the average final spin 
is $a_{\rm spin}=0.07$, for moderate rotators $a_{\rm spin}=0.59$,
while for initially rapidly rotating BHs we obtain $a_{\rm spin}=0.91$. 
The amount of spin up decreases with the initial value of spin, as 
it is more difficult to increase the spin of rapidly rotating objects. 
However, the robust conclusion may be reached in case of BH-NS binaries, 
that independent of initial conditions, it is expected that all BHs are 
spinning. Although accretion can only enhance the spin to a rather limited 
degree, some BHs can spin rapidly provided they were born with high spin. 

The characteristic properties of close BH-NS binaries are illustrated in 
Figure~\ref{Mqt}. Most of the systems host massive BHs with masses $M_{\rm 
bh} \sim 10 \msun$, leading to the rather extreme mass ratio in these 
systems ($q = M_{\rm ns}/M_{\rm bh} \sim 0.14$) as most of neutron stars 
have a mass $M_{\rm ns} \sim 1.3 \msun$.  

The origin of the high mass BHs in our population is due to binary evolutionary 
effects which preferentially select the formation of high mass BHs in these systems. 
Specifically during their evolution most of the BH-NS progenitors proceed without 
any major interaction until the occurrence of the first supernova explosion. The 
more massive primary does not initiate a mass transfer episode, but for most cases 
($98\%$ see {\em acc1} and {\em acc2} channels in Table~2) the less massive 
secondary is involved in such a phase.  
This evolution is related to the high mass ratio of the progenitor systems and 
is connected to the dependence of the rate of mass loss from stellar winds on 
the mass of the star. For example, neutron stars are formed from the stars with 
initial masses of $\sim 10-20 \msun$, while black holes require higher initial 
masses. If the primary is in the low initial mass range of BH formation ($\sim 
20-70 \msun$), then the BH receives a significant kick that tends to disrupt the 
binary. However, if the primary is more massive ($\gtrsim 70 \msun$), BH formation 
proceeds through a direct collapse and does not disrupt the binary.  Such massive 
stars lose their hydrogen rich envelopes very early in their evolution and become 
naked helium stars, leading to the result that they never attain large radii 
($R_{\rm 1,max} \lesssim 200-300 \rsun$). However, the secondary stars (the 
progenitors of NSs) do not lose their hydrogen rich envelopes and they increase 
in size with time. Eventually, during the core-helium burning phase, the stars 
may reach radii ($R_{\rm 1,max} \gtrsim 300-400 \rsun$) sufficient for them to fill 
their Roche lobes, thereby, initiating a CE phase that leads to orbital shrinkage 
and to the formation of the tight BH-NS system. The CE phase, itself, can also
favor the selection of high mass BHs 
since higher mass BHs increase the probability for CE survival due to the greater 
orbital energy available for ejection of the common envelope.
Hence, the high mass BHs found in the BH-NS binaries are 
the combined result of the initial high mass ratio and the common enevelope 
evolution operating on the progenitor systems. 

The tilt distribution of BHs 
is also presented, revealing a drop off with the increasing tilt, 
with about $\sim 50\%$ of the systems characterized by rather low-to-moderate 
tilts ($i_{\rm tilt} < 45^\circ$). Here, we have assumed that both 
supernovae contribute to the tilt of a BH (see eq.~\ref{tilt1}). That is, 
the mass transfer phase between the occurrence of supernovae (found for 
the vast majority of BH-NS progenitors) does not lead to alignment of the 
BH spin with respect to the orbital angular momentum. Had we allowed for such 
the alignment, the results would not change significantly: the distribution
of tilts would look very similar, but with tilts shifted slightly to lower
values (i.e., $\sim 50\%$ of systems with $i_{\rm tilt} < 40^\circ$). Most 
of the BHs ($\sim 65\%$) in BH-NS binaries are formed directly (with no natal 
kick), while the majority of the remaining BHs are formed with a small kick 
(see \S\,2.3). 
Therefore, the first SN explosion (forming BH) does not induce a large
(if any) change of orbital inclination. The tilts mostly originate 
from the second SN explosion, in which the NS is formed (full natal kick). 
Only a small percentage ($\sim 10-15\%$) of systems have tilts that are very 
small $i_{\rm tilt} <5^{\circ}$ independent of the potential alignment 
between the two supernova events. 

Given the results of our BH spin calculations presented in Figures~\ref{aspin}
and ~\ref{Mqt}, in combination with the GRB formation criteria listed in 
Table~\ref{grb}, the fraction ($f_{\rm grb}$) of BH-NS mergers that can 
potentially produce a short-hard GRB can be estimated.  This fraction is 
shown as a function of the initial BH spin in Figure~\ref{grbsel}. For low 
initial BH spins ($a_{\rm spin}<0.5$), only a very small fraction ($f_{\rm 
grb} \sim 1\%$) of BH-NS mergers can potentially produce a GRB, while for 
high initial spins ($a_{\rm spin}>0.6$) the fraction becomes significant 
($f_{\rm grb} \sim 40\%$). The transition occurs for intermediate BH initial 
spins ($a_{\rm spin} \sim 0.55$). In Table~\ref{grb} we have identified the 
three separate GRB formation criteria for BHs in the various mass ranges 
and list the specific fractions of BH-NS systems that satisfy the criteria
in Table~\ref{frac}. The fractions are presented for the three representative 
initial BH spins ($a_{\rm spin}=0.0,\ 0.55,\ 0.9$). The vast majority ($92\%$) 
of BHs fall within pop2 group ($M_{bh}=7-11 \msun$), while only a small fraction 
of BHs ($8\%$) fall within pop1 (see also Fig.~\ref{Mqt}). Note that no BHs 
are formed in close BH-NS binaries with mass over $11 \msun$ (at solar
metallicity) so there are no systems in pop3.  
Consider the dominating population pop2 where the majority 
($86\%$\footnote{Percentages are given in terms of the entire ($100\%$)
close BH-NS population.}) of systems also satisfy the mass ratio criterion 
($0.13<q<0.2$). By imposing the requirement on the BH tilt ($<40^{\circ}$) 
the fraction of potential GRB candidates is reduced to $\sim 0.41$. To determine
the estimated rate for short hard GRBs, a criterion on the final BH spin must 
be imposed.  For this last constraint, we have shown that the final spin 
depends sensitively on the initial BH spin (with a moderate increase from 
binary accretion). Since the spin must be quite large to produce a GRB ($a_{\rm 
spin}>0.6$), systems with low initial spins do not satisfy the spin criterion 
and no GRBs are predicted for pop2 ($f_{\rm grb}=0\%$). If, on the other hand 
moderate-to-large initial spins are assumed, the fraction can be as high as 
$f_{\rm grb}=6-41\%$. The fraction, $f_{\rm grb}=1\%$, for low initial BH 
spins marked in Figure~\ref{grbsel} originate from the small number of BH-NS 
systems fulfilling the criteria in group pop1. Note that these criteria limit 
only the BH mass, mass ratio and tilt, but are independent of BH spin. The 
contribution from group pop1 is small because only a very small fraction of 
BH-NS systems are predicted to host low-mass BHs ($2.5-5 \msun$).

\subsection{BH-BH Binaries}

Similar calculations to the BH-NS binaries were performed for the BH-BH 
binaries, and the accretion histories are listed for BH-BH progenitors
in Table~\ref{bnchan} as well. It is found that most ($70\%$) of these
progenitors do not evolve through a CE nor a RLOF phase after the first BH
was formed. These systems are moderately wide ($\sim 700-900 \rsun$) and
eccentric ($\sim 0.1-0.2$) after the first supernova explosion. The massive
secondary ($\sim 50 \msun$) evolves losing most of its mass in a stellar wind 
finally forming a helium star ($\sim 20 \msun$) that continues to lose mass
until the time of the second supernova explosion ($\sim 10 \msun$). In this case, 
the secondary never fills its Roche lobe. If the secondary collapses to 
form a second BH, such a system would be too wide to coalesce within 10 Gyr. 
However, for a small to a moderate natal kick (see \S\,2.3) associated with the explosion 
the kick may be sufficiently large to induce an extremely high eccentricity for 
some systems ($\sim 0.99$), decreasing the size of the orbit to $\sim 400-500 \rsun$. 
With the two massive black holes ($\sim 6-10 \msun$; see Fig.~10) such a system 
will lose sufficient angular momentum through emission of gravitational radiation 
to coalesce within 10 Gyr (e.g., Peters 1964).

A significant, but small, percentage ($28\%$) of the progenitor 
systems result in the accretion onto the BH during a CE phase. The average 
mass accreted onto the BH ($0.14 \msun$) is slightly higher than for 
BH-NS progenitors ($0.10 \msun$) as BH-BH progenitors are more massive, and a larger mass
reservoir is available during the CE phase. Only a very small percentage 
($2\%$) of systems evolves through a phase of RLOF where accretion occurs 
onto the BH. The distribution of accreted masses is presented in Figure~
\ref{dMaccbb}\footnote{We note that only very few systems are studied 
(a total of 75 close BH-BH binaries) out of which only small fractions 
are encountering CE/RLOF between two supernova events.  Therefore, our 
distributions are burdened with large statistical errors, and are presented
to give an approximate range of the accreted mass rather than a detailed
shape. The number of BH-BH binaries cannot be easily increased as these 
systems are very rare (see Belczynski et al. 2007) and the population 
presented here is the result of the evolution of $40 \times 10^6$ massive 
binaries ($\sim 50\%$ of the Galactic disk); a calculation that consumed 
about 2 months on 50 fast processors.}.

In Figure~\ref{aspinbb} the distribution of BH spins is illustrated for 
BH-BH binaries for several different assumptions on the initial spin. 
Since only the first born BH can evolve through an accretion phase, these
results only apply to these BHs. Only a small fraction (0.3) evolve 
through the accretion phase increasing their spins, while the remaining 
majority and the second born BHs remain at their initial spin.  For the 
initially non spinning case, the average final BHs spin is  
$a_{\rm spin}=0.020$, whereas for moderate and high initial spin the 
final average BH spins are $a_{\rm spin} \sim 0.561$ and $a_{\rm spin} 
\gtrsim 0.904$, respectively. The average BH spin up in BH-BH binaries is 
smaller than for BH-NS systems due to {\em (i)} much smaller fraction of 
accreting BHs in BH-BH binaries, and {\em (ii)} the fact that BHs in BH-BH
binaries accrete mostly through CE, while significant fraction of BHs in 
BH-NS binaries accrete additionally (to CE) in RLOF thus enhancing their 
spin up. Note also, that the highest attainable spins for BHs in BH-BH
binaries are close to their initial spins.   

The physical properties of merging BH-BH binaries are shown in Figure~
\ref{Mqtbb}. In the upper panel, the mass distribution for the first born 
and second born BH are displayed separately. 
Average mass of the first born BH ($\sim 6.7 \msun$) is smaller than the 
second born BH ($\sim 8.3 \msun$). This is a result of the initial mass transfer 
phase encountered for many BH-BH progenitors. This mass transfer occurs when
the secondary is still on main sequence, and the evolved primary transfers 
mass and rejuvenates the secondary. Rejuvenation leads to an increase of the
secondary core mass and eventually leads to a higher BH mass if sufficient 
matter is transferred. Most of the BHs are massive ($M_{\rm bh} \gtrsim 6 
\msun$), however there are a number of systems with lower mass BHs.  
Note that there is no similar high mass BH selection effect as for BH-NS
binaries (see \S\,3.1) since the majority of BH-BH progenitor stars lose
their hydrogen envelopes and do not expand sufficiently to initiate a mass
transfer episode (see {\em acc5} formation channel in Table~2). Hence, the resulting 
mass ratio distribution is skewed toward comparable mass BHs, unlike for 
case of BH-NS systems. These distributions are slightly different than
presented earlier (e.g., Belczynski et al. 2007) due to the changes in
treatment of rejuvenation (see \S\,2.1).
The new results are presented for tilt distribution where it is found 
that most of the tilts are insignificant (i.e., $50\%$ of systems 
have tilts $i_{\rm tilt} < 1^\circ$), with a long tail distribution 
reaching high values ($i_{\rm tilt} \sim 100^\circ$). The small tilts are
the result of the direct BH formation (no kick assumed) that occurs for most 
massive BHs. Since most BHs in close BH-BH binaries are massive the tilts are 
very small. In the few cases in which tilts are significant they arise from 
both SNe since for most BH-BH systems there is no mass transfer phase
(no potential alignment) between the two supernovae.

\subsection{Comparison to Earlier Work}

The results of our study differ from the recent work of O'Shaughnessy et 
al. (2005), also based on the {\tt StarTrack} code, and reflect the 
introduction of additional input physics and a different implementation 
of the code. In particular, O'Shaughnessy et al. (2005) find that the 
masses of BHs in BH-NS binaries range from $2-15 \msun$ and 
that the accretion of mass onto BHs can amount to as much as $\sim 5 \msun$ 
(see their Fig.1).  This leads to a significant increase in the BH spin 
independent of the initial BH spin (see their Fig.2 and 3). We note that 
the majority of the BHs in BH-NS binaries in O'Shaughnessy et al. (2005) are 
of low mass, starting as heavy NSs ($\lesssim 2 \msun$) that accreted sufficient
mass to exceed the maximum NS mass limit (adopted to be $2 \msun$). In addition, 
the full Bondi-Hoyle accretion onto the compact objects (both NS and BH) in the 
CE phase was adopted. In contrast, a higher NS mass limit ($2.5 \msun$) is 
adopted here, guided by the recent NS mass estimates (e.g., $\sim 2.7 \msun$ 
pulsar mass of PSR J1748-2021B in NGC 6440, Freire et al. 2007, although this
result needs still to be confirmed), and only modest 
mass accretion takes place during the CE phase (see below). The combination of 
these two effects depletes O'Shaughnessy et al. (2005) BH-NS population by 
factors of $\sim 4-5$ leaving most systems with high mass BH (i.e., extreme 
mass ratios, as observed in our current study). The highest amount of accretion 
onto a BH in the CE phase found here is $\sim 0.2 \msun$ (see Fig.~\ref{dMacc}
or Fig.~\ref{dMaccbb}), 
while it reaches $\sim 5 \msun$ in O'Shaughnessy et al. (2005). This large 
difference stems from the combination of the following.  Only 10\% of 
the accretion in the CE phase is assumed, guided by the hydrodynamical simulations 
of Ruffert (1999) and recent estimates of Ricker \& Taam (2008) and noting that only 
a minimal accretion is permitted in the CE phase if one is to reproduce the observed 
NS mass spectrum (Belczynski et al. 2008b). In addition, we have only considered a CE 
model with a high efficiency of envelope ejection ($\alpha_{\rm ce}=1$; e.g., see 
Webbink 1984) while O'Shaughnessy et al.  (2005) use an entire spectrum of efficiencies 
($\alpha_{\rm ce}=0-1$). It is to be noted that for typical BH-NS progenitors 
evolving through the CE phase, a change of the efficiency from our adopted 
value ($\alpha_{\rm ce}=1$) to much smaller values of 0.1 and 0.01 leads to a 
factor of 2 and 4 increase in the amount of accreted mass onto the BH 
respectively. This is due to a fact that for smaller CE efficiencies the compact 
object sinks further into the donors envelope, resulting in greater 
accretion of mass, to supply sufficient energy for the ejection of the envelope. 
We note that very small CE efficiency values ($\alpha_{\rm ce} \sim 0.01$) 
tend to eliminate (through CE mergers) most of NS-NS progenitors, and lower the 
NS-NS predicted merger rates below the empirically estimated rates. Finally, 
O'Shaughnessy et al. (2005) allow for a 
wide range of input parameters (such as decreased stellar winds, or fully
conservative mass transfer) that eventually lead to more massive donors at 
the time of the CE phase. This provides a larger mass reservoir for accretion
onto the BH, however, this effect is less significant than the other two effects
mentioned above.

\section{Discussion}

The evolution of the progenitors of close (coalescing) BH-NS and BH-BH 
binaries has been investigated with particular emphasis  on the spin 
and tilt of the BH members in these systems. The results of our population 
synthesis have been used to estimate the fraction of BH-NS mergers that 
can potentially produce a short-hard GRB within the context of the coalescing 
binary model for these GRBs. It is found that close BH-NS systems form with 
rather massive BH ($\sim 10 \msun$), resulting in a mass ratios typically of 
the order of $0.14$.  Accretion onto BHs in the 
progenitors of close BH-NS binaries leads to a small increase in the BH 
spin.  Thus, the final BH spin is only a weak function of binary accretion 
and primarily depends on its unknown initial value. The 
misalignment of the BH spin with respect to the direction of the binary 
angular momentum is found to be moderate ($\lesssim 45^\circ$ for $50\%$ 
of systems) and is mainly induced by the second SN explosion that forms 
the NS in the system. 

By combining our results with the recent hydrodynamical calculations of BH-NS
mergers, we estimate the fraction of these mergers which can potentially 
produce a short-hard GRB. It is found that only a very small percentage 
($\sim 1\%$) of BH-NS mergers can produce GRB if the initial BH spins are 
small ($a_{\rm spin} <0.5$). This percentage increases to $\sim 40\%$ if the 
initial BH spins are high ($a_{\rm spin}>0.6$). 
We point out that an estimate of a GRB rate originating from BH-NS mergers 
should take into account the reduction factor of the order of $\sim 2.5-100$ 
compared to previous estimates. Given this reduction factor to the already 
low BH-NS merger rate the observed short-hard GRBs are difficult to reconcile 
with the BH-NS merger scenario only. 
 
In our calculations we have used a uniform distribution for the direction of the
natal kicks. However, some recent studies have pointed out that there may exist
NS spin -- kick velocity correlation that would favor polar kicks (along
spin axis) over uniformly distributed kicks (e.g., Rankin 2007; Willems et
al. 2008; Postnov \& Kuranov 2008). The effect of polar kicks on BH-NS
progenitors is two fold: {\em (i)} it reduces the number of BH-NS systems as
polar kicks tend to disrupt binaries more easily than uniform kicks, {\em (ii)} 
the tilt angle (on average) is larger for surviving binaries, as the NS is
kicked straight out of its orbital plane.
Both of the above decrease the number of potential BH-NS GRB
progenitors, further supporting our conclusion that these systems are
unlikely to be responsible for the majority of short-hard GRBs.

\begin{figure}
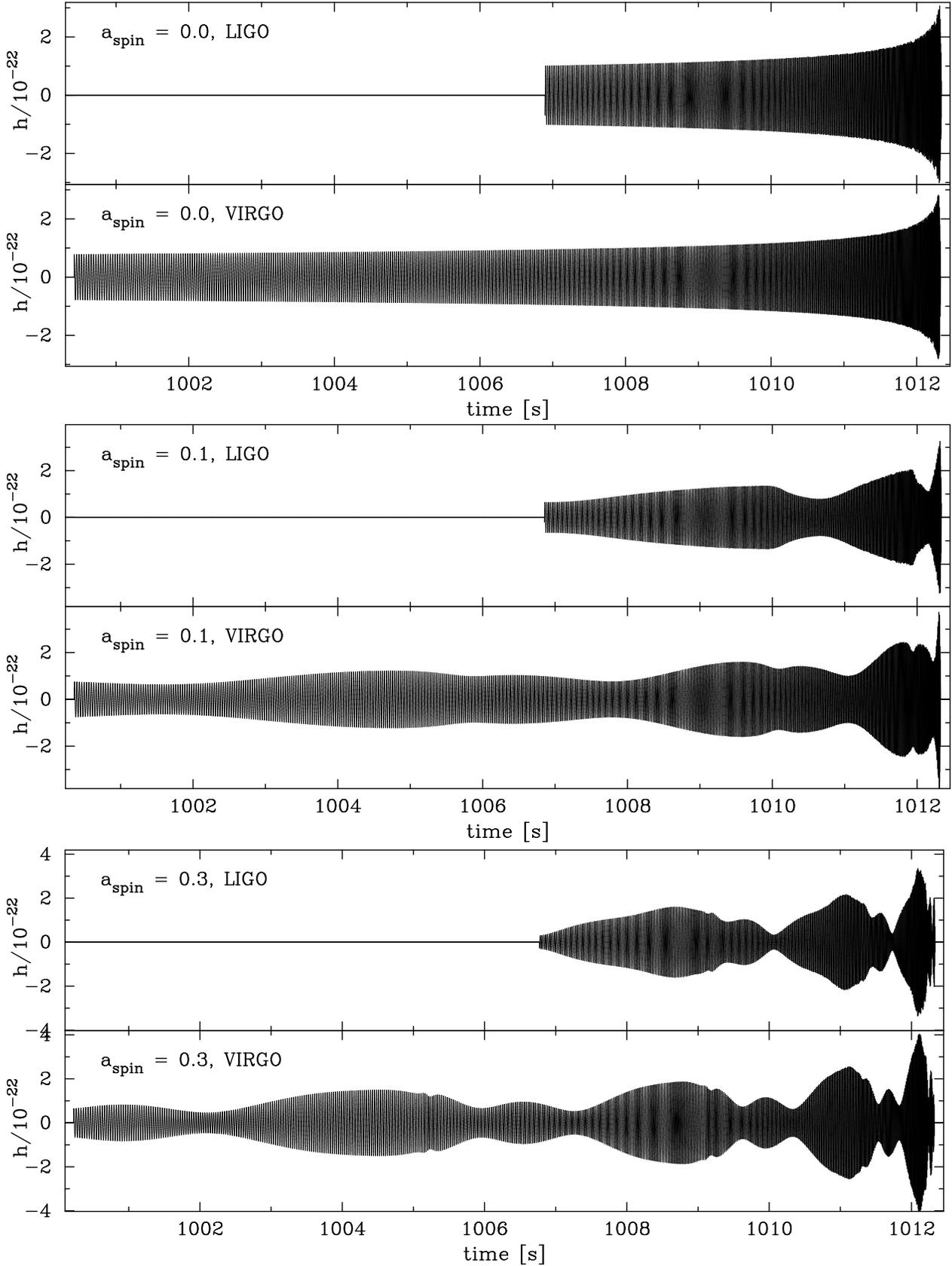

\includegraphics[width=0.4\columnwidth,angle=270]{f3a.ps}\\
\includegraphics[width=0.4\columnwidth,angle=270]{f3b.ps}\\
\includegraphics[width=0.4\columnwidth,angle=270]{f3c.ps}
\caption{
Gravitational radiation inspiral signal ($h$ -- GR wave strain) of BH-NS binary 
with $10 \msun$ BH and $1.4 \msun$ NS at a distance of 30 Mpc. The signal was 
calculated for LIGO and VIRGO detectors for non-spinning case (unrealistic)  
and low-spin cases with $a_{\rm spin}=0.1, 0.3$ and moderate tilt of 
$i_{\rm tilt} = 35^\circ$. Note that as compared with a non spinning case there is 
only a small difference in signal for $a_{\rm spin}=0.1$, and that the
difference becomes more pronounced at higher spins $a_{\rm spin}=0.3$. 
However, only a very few black holes ($\sim 1\%$) in BH-NS systems accrete enough 
to be spun up to attain spins above $a_{\rm spin}=0.3$ (provided that black holes 
were initially non-spinning; see top panel of Fig.~5). 
}
\label{grsig}
\end{figure}

\begin{figure}
\includegraphics[width=1.0\columnwidth]{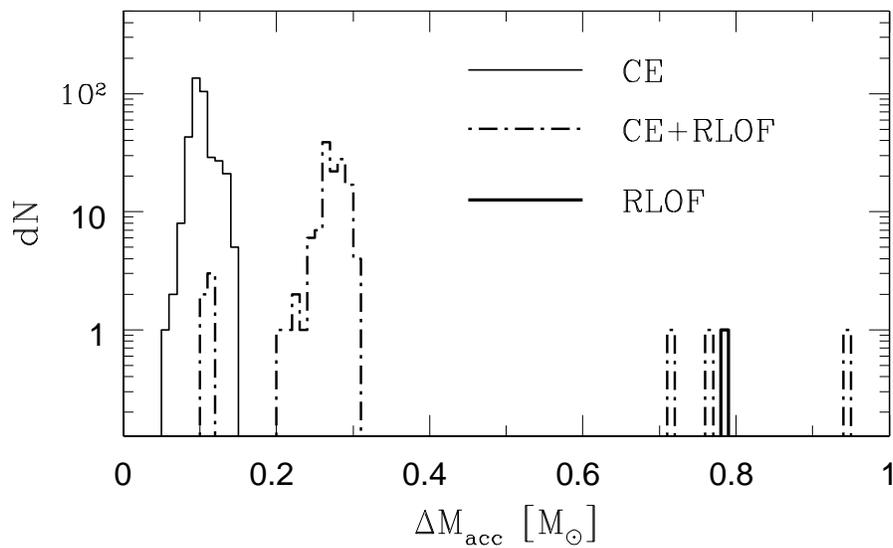}
\caption{
Mass accreted onto the BH during the evolutionary history leading to the 
formation of close BH-NS binaries. Mass can be accreted during the common 
envelope (CE), stable Roche lobe overflow (RLOF), or during a combination 
of the above modes. Note that BHs do not accrete a significant amount of mass
($\lesssim 0.3 \msun$) throughout their evolution. In this calculation BHs were 
assumed to be born with moderate initial spin ($a_{\rm spin}=0.55$). 
}
\label{dMacc}
\end{figure}

\begin{figure}
\includegraphics[width=1.0\columnwidth]{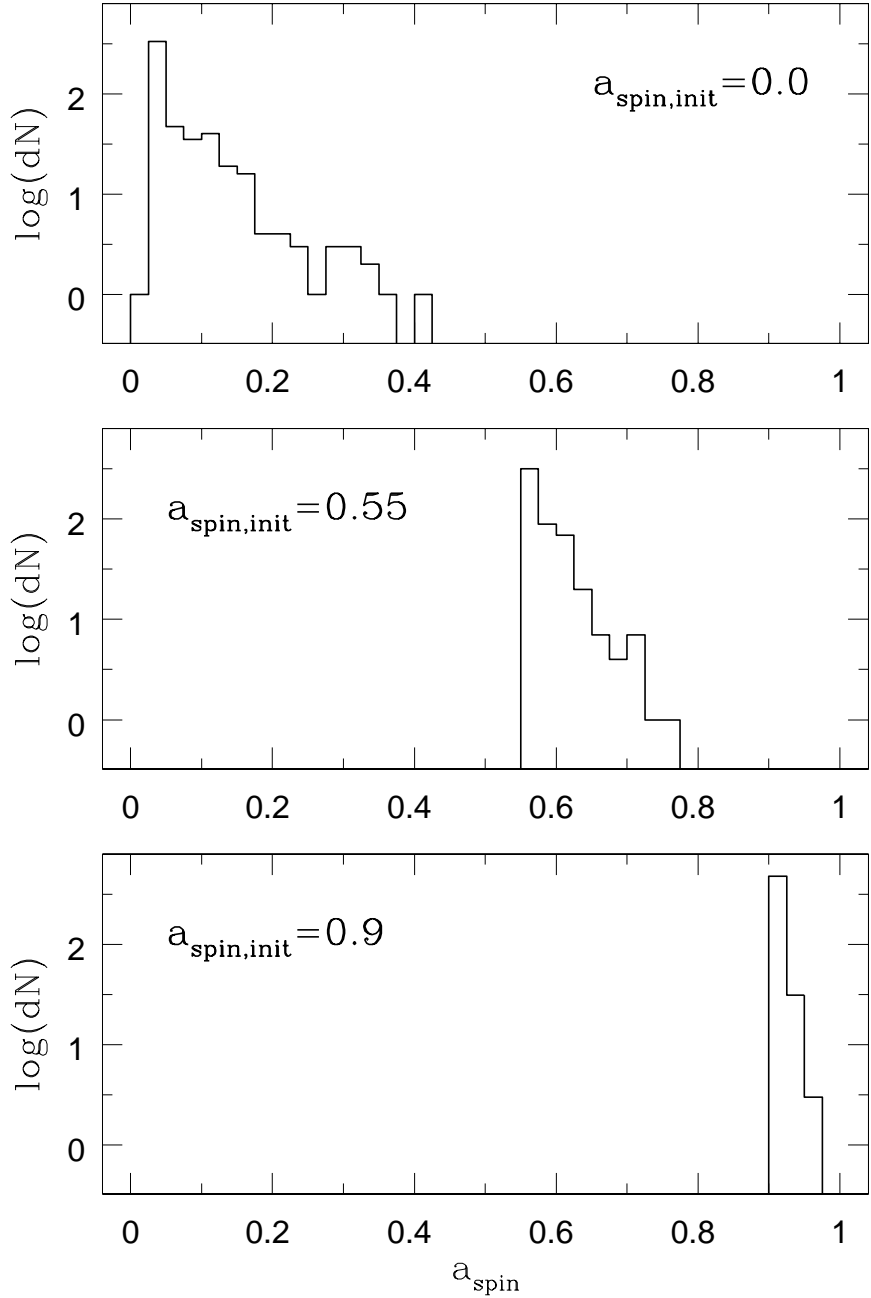}
\caption{
Distribution of BH spins in close BH-NS binaries. Each calculation was
performed with a different initial BH spin. Note that for the case of
initially non spinning BHs (top panel), there is a small increase of BH 
spin due to binary accretion: $\sim 20\%$ of BHs increase their spin above 
$a_{\rm spin}=0.1$, while only $\sim 1\%$ attain spins above 
$a_{\rm spin}=0.3$. However, high spins ($a_{\rm spin} \geq 0.9$) can be 
obtained only if BHs are initially formed with high spins (bottom panel). 
}
\label{aspin}
\end{figure}

\begin{figure}
\includegraphics[width=1.0\columnwidth]{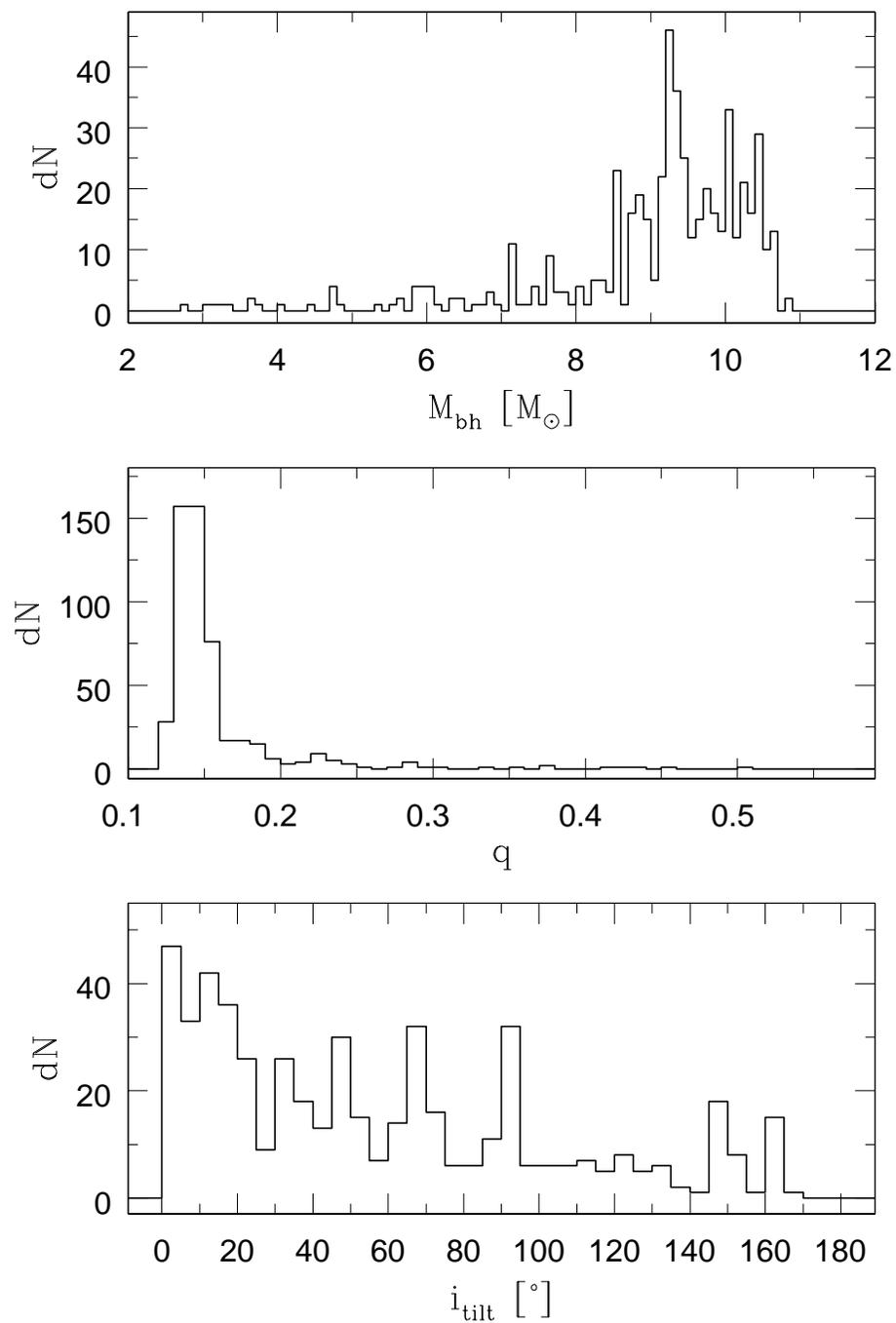}
\caption{
Distribution of BH mass, mass ratio and tilt of BH spin in close BH-NS 
binaries for a model with the moderate initial BH spin ($a_{\rm spin}=
0.55$). Note that the high masses of BHs result in an extreme mass 
ratio distribution and that moderate tilts dominate ($i_{\rm tilt} < 
40^\circ$ for $\sim 50\%$ of systems).}
\label{Mqt}
\end{figure}

\begin{figure}
\includegraphics[width=1.0\columnwidth]{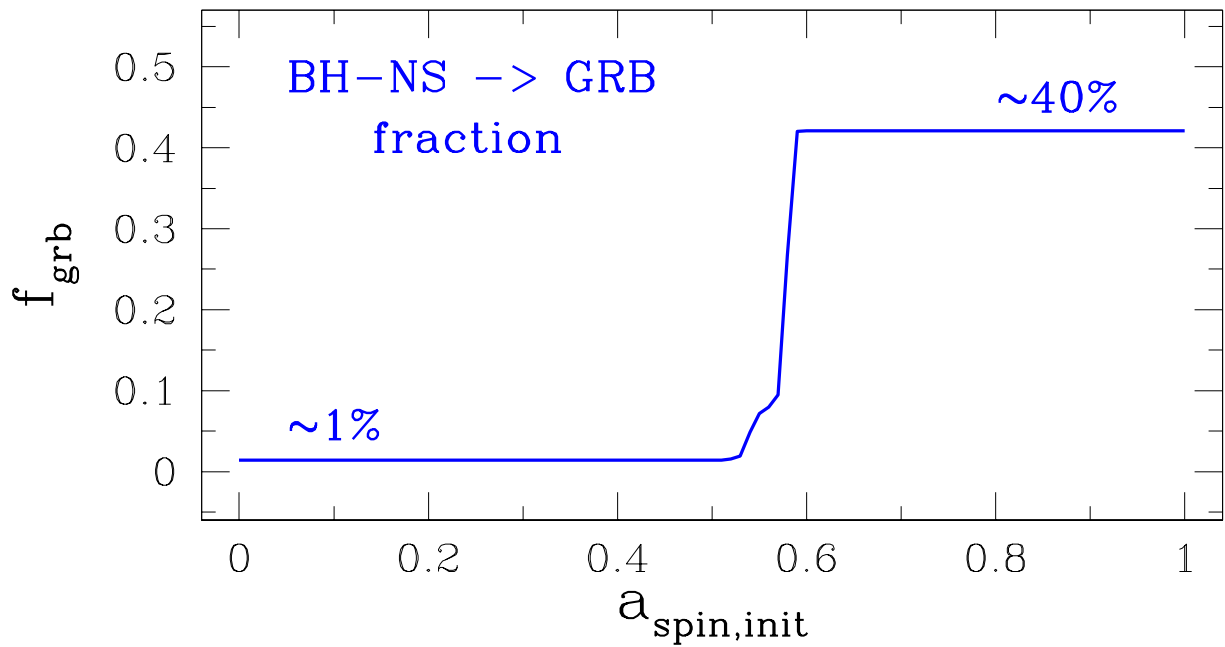}
\caption{
Fraction of BH-NS mergers that can produce a short-hard GRB according to the 
criteria presented in \S\,2.4. Note the strong dependence
on the assumed initial BH spin; only a very small fraction ($\sim 1\%$) of
the mergers can produce GRB for low initial spins, while significant
fraction ($\sim 40\%$) is found for high initial BH spins.  
}
\label{grbsel}
\end{figure}

\begin{figure}
\includegraphics[width=1.0\columnwidth]{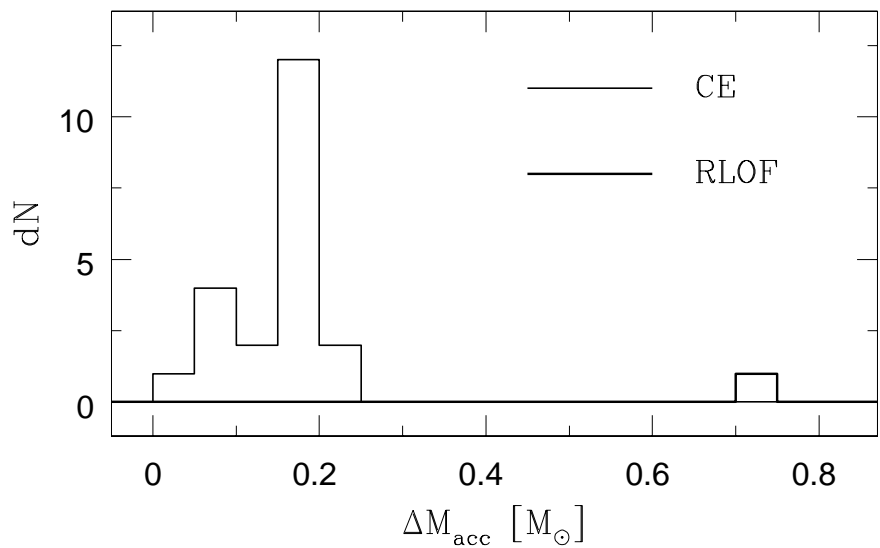}
\caption{
Mass accreted onto BHs during the evolutionary history leading to formation of
close BH-BH binaries for a model in which BHs were assumed to be born with 
a moderate initial spin ($a_{\rm spin}=0.55$). Mass can be accreted either 
during the common envelope (CE), or stable Roche lobe overflow (RLOF) phase.}
\label{dMaccbb}
\end{figure}

\begin{figure}
\includegraphics[width=1.0\columnwidth]{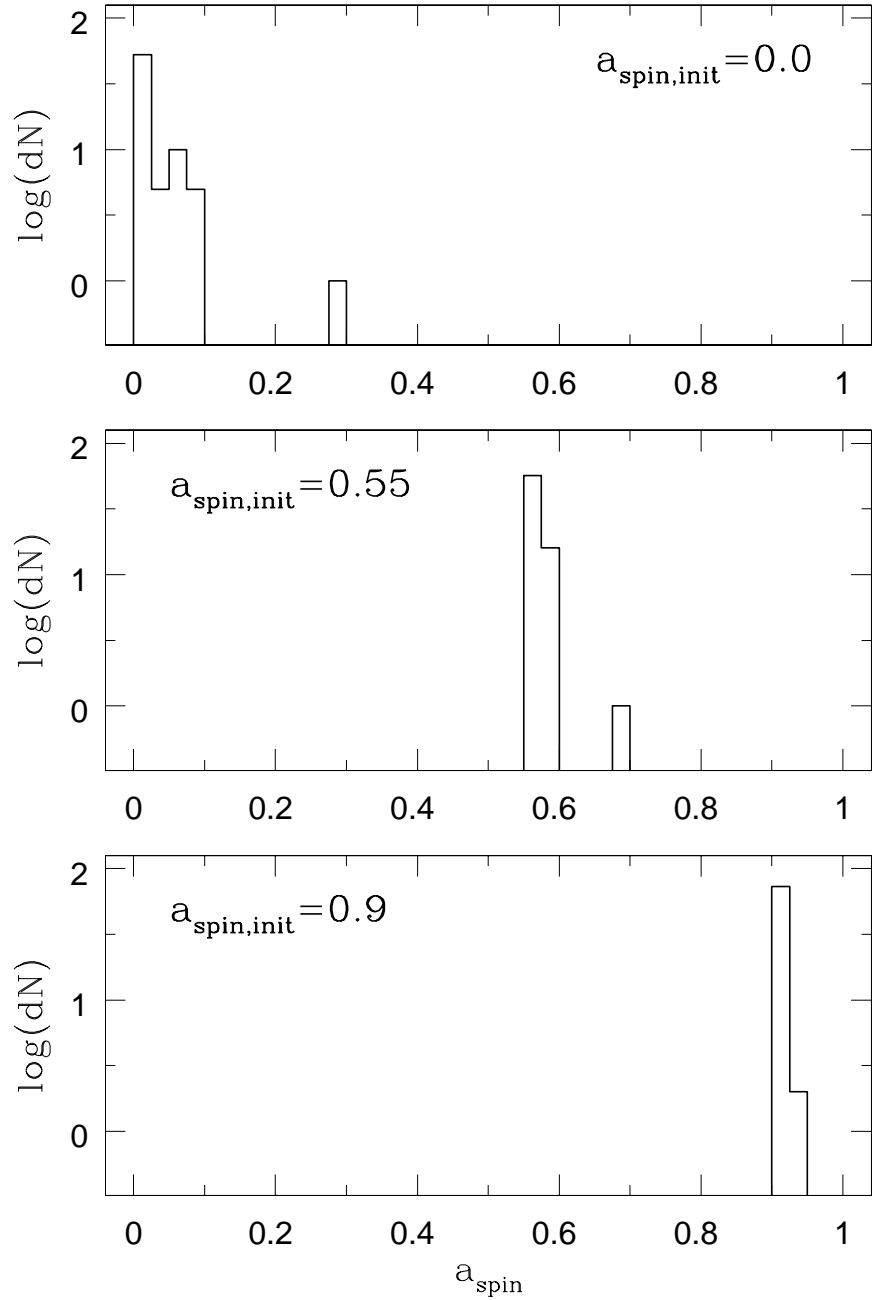}
\caption{
Distribution of BH spins in close BH-BH binaries for models characterized by 
an initial spin given as $a_{\rm spin}=0, 0.55$ and 0.9 for the upper, middle 
and lower panel respectively.}
\label{aspinbb}
\end{figure}

\begin{figure}
\includegraphics[width=1.0\columnwidth]{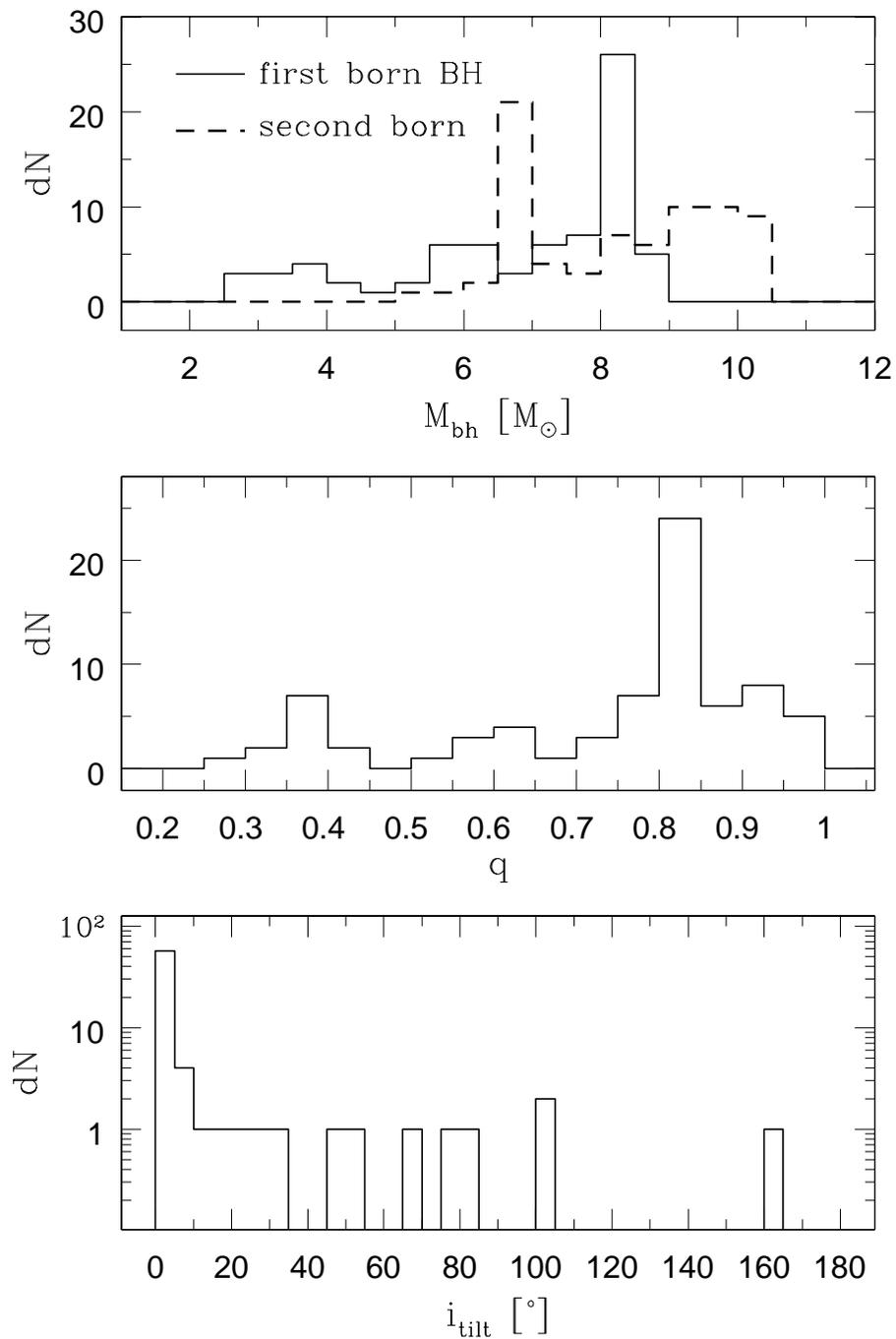}
\caption{
Distribution of BH mass (upper panel), mass ratio (middle panel), and tilt of 
BH spin (lower panel) in close BH-BH binaries for a model with a moderate 
initial BH spin ($a_{\rm spin}=0.55$). The mass distribution of the first born 
BH and the second born BH are shown separately.}
\label{Mqtbb}
\end{figure}

The spin evolution of BHs in close BH-BH binaries was also investigated. In  
this case, only a fraction (0.3) of BHs in these systems slightly  
increase their spins due to an accretion phase in the binary progenitor. This 
small fraction is due to the fact that many BH-BH systems do not experience 
a mass transfer episode after the first BH is formed. As described in the 
previous section, almost all BHs in close BH-NS binaries increase their spin  
(even for initially non-spinning BHs, the average final BH spin is found to
be $a_{\rm spin} \sim  0.1$). However, the change of spin is rather insignificant
and does not alter drastically the shape of the GR inspiral signal. As illustrated 
in Figure~\ref{grsig} a small spin magnitude ($\sim 0.1$) with a moderate tilt to 
the orbital angular momentum vector does has a large effect on the inspiral 
waveform. Only for higher spins ($\gtrsim 0.3$) can the waveform be significantly
altered. However, only a very small fraction ($\sim 1\%$) of initially non spinning 
BHs can attain such spins (see Fig.~5). 

Currently (Abbott et al. 2005; Abbott et al. 2006), non spinning templates are 
used in the search for an inspiral signal in LIGO data streams. Grandclement et al. 
(2004) estimated that the loss of BH-NS inspiral detection due to use of non 
spinning templates can not be greater than $\sim 30\%$. As this estimate employed 
the possibility that all BHs in BH-NS binaries may have misaligned spins,
and as we have demonstrated that {\em (i)} misalignment is rather moderate
and {\em (ii)} accretion does not significantly increase BH spin, this
loss estimate should be treated as an upper limit. 
For BH-BH binaries the potential detection loss is expected to be even smaller, 
as most of these binaries are predicted to have BH spins aligned with the orbital 
spin, and the accretion is insignificant in BH spin evolution. 
Although the detection of BH-NS and BH-BH binaries is 
not significantly affected by the use of non spinning templates, the parameter 
estimation for a detected system should incorporate techniques that allow 
for spinning and misaligned BHs. Our estimates of the BH spin can be used as 
a guide for the initial conditions in hydrodynamical and detailed relativistic 
simulations of the BH-BH and BH-NS mergers and their expected gravitational-wave 
signature. In addition, BH spin is an important parameter required in the estimation of the 
gravitational radiation recoil produced by the merger of two spinning black 
holes (see Baker et al. 2007 and references therein).

Finally, we point out that the measurement of BH spin for the inspiraling BH 
binary, can yield a value of initial BH spin, because the majority of BH-BH 
binaries are mostly unaffected by accretion spin up. This may be the 
most direct way to infer the initial BH spin.  We note, however, that BH-BH 
binaries are predicted to be very rare in field populations and may be difficult 
to observe as only $\sim 2$ detections per year for advanced LIGO (Belczynski et 
al. 2007) are expected. If significantly greater numbers of BH-BH binaries are
detected, then they most probably originate from dynamical formation in
globular clusters (Sadowski et al. 2008). In this case a large misalignment 
(formation through tidal capture/exchange) for BH-BH binaries is possible.

\acknowledgements
We wish to thank the anonymous referee for very useful comments. 
We also express special thanks to Sophia Cisneros, Richard O'Shaughnessy, Alberto
Vecchio, Vicky Kalogera, Fred Rasio, Joan Centrella and John G. Baker for 
useful discussions and comments. KB thanks Northwestern University Theory 
Group for hospitality. We acknowledge partial support through KBN Grant 
1P03D02228 and 1P03D00530 (KB), NSF Grant Nos. AST-0200876 and AST-0703960 (RT), 
and by the Theoretical Institute for Advanced Research in Astrophysics (TIARA) 
operated under Academia Sinica and the National Science Council Excellence
Projects program in Taiwan administered through grant number NSC
96-2752-M-007-007-PAE.

\clearpage

\begin{deluxetable}{ccccc}
\tablewidth{300pt}
\tablecaption{GRB Formation Criteria\tablenotemark{a}}
\tablehead{ Name & $M_{\rm bh}$ [$\msun$] & q & $i_{\rm tilt}$ [$^\circ$] & $a_{\rm spin}$} 
\startdata

pop1                & 2.5-7   & 0.35-0.7  & $< 90$ & $> 0$   \\
pop2                & 7-11    & 0.13-0.2  & $< 40$ & $> 0.6$ \\
pop3                & 11-15   & 0.09-0.12 & $< 40$ & $> 0.9$ \\
\enddata
\label{grb}
\tablenotetext{a}{Detailed description of criteria is given in \S\,2.4.}
\end{deluxetable}

\begin{deluxetable}{lrlc}
\tablewidth{350pt}
\tablecaption{Accretion History for BH Binary Progenitors}
\tablehead{ Name & Efficiency & Formation History\tablenotemark{a} & Mass
Accreted\tablenotemark{b}}
\startdata

acc1 (BH-NS) & 73\% & ..... SN1 CE SN2 & $0.10 \msun$ \\
acc2 (BH-NS) & 26\% & ..... SN1 CE RLOF SN2 & $0.28 \msun$ \\
acc3 (BH-NS) & 0.5\% & ..... SN1 RLOF SN2 & $0.79 \msun$ \\
acc4 (BH-NS) & 0.5\% & ..... SN1 SN2 & no accretion \\
                    &                    &        \\
acc5 (BH-BH) & 70\% & ..... SN1 SN2 & no accretion \\
acc6 (BH-BH) & 28\% & ..... SN1 CE SN2 & $0.14 \msun$ \\  
acc7 (BH-BH) &  2\% & ..... SN1 RLOF SN2 & $0.73 \msun$ \\

\enddata
\label{bnchan}
\tablenotetext{a}{The evolutionary history after the first supernova
(SN) that forms a BH is listed. Mass accretion may occur in the common 
envelope (CE) and/or during stable Roche lobe overflow (RLOF) phase. }
\tablenotetext{b}{Average accreted mass listed. For the full distribution
see Fig.~\ref{dMacc} and Fig.~\ref{dMaccbb}.} 
\end{deluxetable}

\begin{deluxetable}{ccccc}
\tablewidth{300pt}
\tablecaption{GRB Formation Fractions\tablenotemark{a}}
\tablehead{Criterion:& $M_{\rm bh}$ & q & $i_{\rm tilt}$ & $a_{\rm spin}$}
\startdata

$a_{\rm spin}=0$:    &        &           &        &           \\
pop1                 &  0.08  &  0.02     & 0.01   &  0.01     \\
pop2                 &  0.92  &  0.86     & 0.41   &     0     \\
pop3                 &     0  &     0     &    0   &     0     \\

$a_{\rm spin}=0.55$: &        &           &        &           \\
pop1                 &  0.08  &  0.02     & 0.01   &  0.01     \\
pop2                 &  0.92  &  0.86     & 0.41   &  0.06     \\
pop3                 &     0  &     0     &    0   &     0     \\ 

$a_{\rm spin}=0.9$:  &        &           &        &           \\
pop1                 &  0.08  &  0.02     & 0.01   &   0.01    \\
pop2                 &  0.92  &  0.86     & 0.41   &   0.41    \\
pop3                 &     0  &     0     &    0   &      0    \\

\enddata
\label{frac}
\tablenotetext{a}{The fraction of BH-NS mergers that satisfy the GRB 
criteria presented in Table~\ref{grb} are listed. Imposition of the 
subsequent criteria (on BH mass, mass ratio, tilt and spin magnitude) 
reduces the total fraction.}
\end{deluxetable}

\clearpage

\clearpage

\clearpage

\clearpage

\clearpage

\clearpage

\clearpage

\clearpage

\clearpage

\clearpage

\end{document}